\documentclass[]{elsart}

\usepackage{epsfig}
\usepackage{graphics}
\usepackage{graphicx}
\usepackage[centertags]{amsmath}
\usepackage{multicol}
\usepackage{url}
\usepackage{subfigure}
\usepackage{algorithmic}
\usepackage{fancyhdr}

\newtheorem{proposition}{Proposition}
\newtheorem{finding}{Finding}

\begin{document}


\title{Simulating  strange attraction of acellular slime mould \emph{Physarum polycephalum} to herbal tablets }

\author{Andrew Adamatzky} 

\address{University of the West of England, Bristol BS16 1QY, United Kingdom \\ 
{\tt andrew.adamatzky@uwe.ac.uk}\\}

\vspace{0.5cm}

{\small   Final edited version of this paper is published in\\ 
Andrew Adamatzky: Simulating strange attraction of acellular slime mould \emph{Physarum polycephalum} to herbal tablets. 
\emph{Mathematical and Computer Modelling} 55(3-4): 884-900 (2012)
}

\date{}

\maketitle

\begin{abstract}

\noindent
Plasmodium of acellular slime mould \emph{Physarum polycephalum} exhibits traits of wave-like behaviour. The plasmodium's behaviour can be finely tuned  in laboratory experiments by using herbal tablets. A single tablet acts as a fixed attractor: plasmodium propagates towards the tablet, envelops the tablet with its body and stays around the tablet for several days. Being presented with several tablets the plasmodium executes limit cycle like motions. The plasmodium performs sophisticated routines of movement around tablets: rotation, splitting, and annihilation.  We use to two-variable Oregonator model to simulate the plasmodium behaviour in presence of the herbal tablets.  Numerical experiments confirm that using long-distance attracting and short-distance repelling fields we can  organise  arbitrary movement of plasmodia.

\vspace{0.5cm}

\end{abstract}


\pagestyle{fancy}


\newpage

 \lhead{{\footnotesize Adamatzky~A. Mathematical and Computer Modelling 55 (2012) 884--900.}}
 \rhead{.}
 \chead{.}

 \small
 
\section{Introduction}

Plasmodium is a vegetative stage of acellular slime mould \emph{Physarum polycephalum}~\cite{stephenson_2000}. The plasmodium  is a syncytium, a single cell with  many nuclei, which feeds on microscopic particles.
During its foraging behaviour the plasmodium spans scattered sources of nutrients with a network of  protoplasmic tubes. The protoplasmic network is optimised to cover all sources of food and to provide a robust and speedy transportation of nutrients and metabolites in the plasmodium body. The plasmodium's foraging behaviour can be interpreted as computation: data are represented by spatial configurations of attractants and repellents, and  results of computation by structures of protoplasmic network formed by the plasmodium on the data sets~\cite{nakagaki_2000,nakagaki_2001a,adamatzky_physarummachines}. The problems solved by plasmodium of \emph{P. polycephalum} include shortest path~\cite{nakagaki_2000,nakagaki_2001a}, 
implementatiton of storage modification machines~\cite{adamatzky_ppl_2007},
Voronoi diagram~\cite{shirakawa},  Delaunay triangulation~\cite{adamatzky_physarummachines}, 
logical computing~\cite{tsuda_2004,adamatzky_gates}, and 
process algebra~\cite{schumann_adamatzky_2009}; 
see overview in~\cite{adamatzky_physarummachines}.

Slime mould based computers are programmed using attractants and repellents. While reaction of \emph{P. polycephalum} to repellents, e.g. illumination-~\cite{nakagaki_yamada_1999}, thermo-~\cite{tso_mansour_1975, matsumoto_1980} and salt-based repellents~\cite{adamatzky_physarum_salt}, is quite straightforward the plasmodium's behaviour in presence of chemo-attractants still remains a challenging topic of research~\cite{dussutour_2010}.

Studies on attractants of \emph{P. polycephalum} can be traced back to early 1900s, with interest usually reignited every 20-30 years, see overview of pre-1960s works in~\cite{carlile_1970}.  A common logic behind explaining attraction of plasmodium to various substances was that slime mould prefers substances with potentially high nutritional value, particularly those high on carbohydrates. Laboratory experiments in 1970s demonstrated that plasmodium is attracted to glucose, maltose, mannose and galactose~\cite{carlile_1970,knowles_carlile_1978}. Recently there was a renewed interest in studying relations between chemo-attractants and their nutritional values~\cite{dussutour_2010}. Molecular mechanisms of carbohydrate recognition  are proposed in~\cite{kouno_2011}. 

Already in late 1970s it became clear that nutritional value is not the only prerequisite for a  substance to be a chemo-attractant for \emph{P. polycephalum}~\cite{kincaid_mansour_1978, kincaid_mansour_1978a}. It was experimentally proved that the slime mould is attracted by peptones~\cite{coman_1940,carlile_1970}, aminoacids phenylalanine, leucine, serine, asparagine, glycine, alanine, aspartate, glutamate, and threonine~\cite{chet_1977,kincaid_mansour_1978,mcclory_coote_1985}. There are reports that plasmodium is indifferent to sucrose, fructose and ribose~\cite{carlile_1970, knowles_carlile_1978}, 
and repelled by sucrose~\cite{ueda_1976}  and tryptophan~\cite{mcclory_coote_1985}. Moreover, some of the 
chemo-attractants may inhibit plasmodium's movement when the plasmodium comes into direct contact with them, e.g.
galactose and mannose attract plasmodium~\cite{carlile_1970,knowles_carlile_1978} 
but also inhibit plasmodium's motion~\cite{denbo_miller_1978}. 

While perfecting our concept and experimental implementations of Physarum machines ---  programmable amorphous biological computers experimentally implemented in plasmodium of \emph{P.polycephalum}~\cite{adamatzky_physarummachines} we  tried to find a universal way of programming plasmodium activity. A universal programming substance is the on that can be used to immobilise the plasmodium when necessary but also set the plasmodium in a complex pattern of motion when required. We found that the herbal tablets with alleged sleep-inducing properties --- Nytol, Kalms Sleep and Kalms Tablets --- are universal programming substances. The paper presents  outcomes of our studies and illustrates basic routines of programming the plasmodium's behaviour.

\section{Laboratory methods}
\label{experimental}

\begin{figure}[!tbp]
\centering
\subfigure{\includegraphics[width=0.49\textwidth]{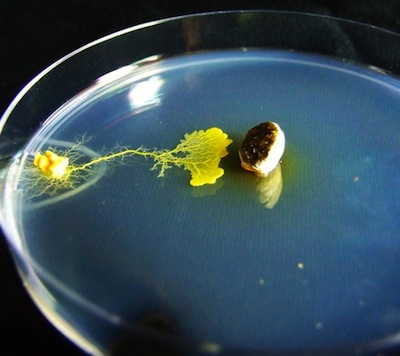}}
\subfigure{\includegraphics[width=0.49\textwidth]{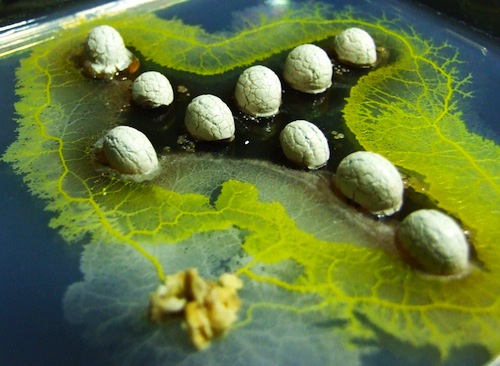}}
\caption{Photographs of a typical experimental setups. 
(a)~Plasmodium propagates towards Kalms Sleep tablet.
(b)~Plasmodium envelopes a compact configuration of Kalms Sleep tablets.}
\label{photoexperiment}
\end{figure}

We cultivated \emph{P. polycephalum} in plastic containers, on paper towels sprinkled with
still drinking water and fed with rolled oats. For experiments we used polystyrene Petri dishes (round, diameter
120~mm, and rectangular $120 \times 120$~mm), and 2\% agar gel (Select agar, Sigma Aldrich) as a 
non-nutrient substrate and 2\% Corn Meal Agar (Fluka Analytical, Sigma Aldrich) as a nutrient-rich substrate.   

We represented attracting domains with three types of sleep-inducing herbal tablets:  Kalms Tablets and Kalms Sleep
(G.~R.~Lane Health Products Ltd, Gloucester GL2 0GR, UK) and Nytol (Brunel Healthcare Manufacturing Ltd, Derbyshire, 
DE11 0BB, UK). We placed an oat flake colonised by plasmodium of 
\emph{P. polycephalum} at the fringe of a Petri dish and distributed few half-pills of Kalms Tablets, Kalms Sleep or Nytol in the central area of the dish (Fig.~\ref{photoexperiment}).  Images of plasmodium are recorded by scanning Petri dishes in Epson Perfection 4490 (thus most images are views of Petri dishes from below). Photos are taken using FujiPix 6000 camera.

\section{Laboratory experimental results}
\label{experimentalresults}

\begin{figure}[!tbp]
\centering
\subfigure[]{\includegraphics[width=0.49\textwidth]{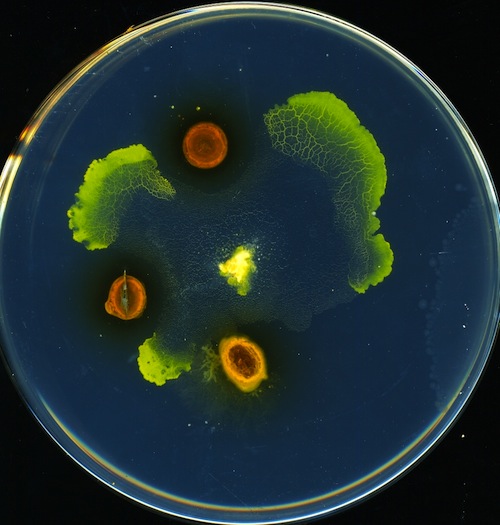}}
\subfigure[]{\includegraphics[width=0.49\textwidth]{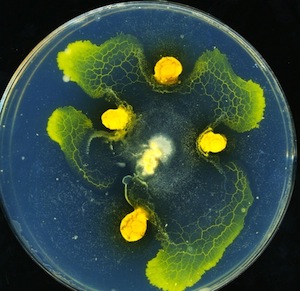}}
\caption{Snapshots of plasmodium propagating on a nutrient substrate with three Kalms Sleep~(a) and Kalms Tablets~(b) tablets. Snapshots are taken 20~h after inoculation of the plasmodium in the centre of a Petri dish.}
\label{nutrientsubstrate}
\end{figure} 

\begin{finding}
Herbal sleep-remedy tablets are treated by plasmodium of P. polycephalum as obstacles when  the plasmodium 
grows on a nutrient-substrate.
\end{finding}

Being inoculated on a nutrient substrate plasmodium of \emph{P. polycephalum} propagates omni-directionally, in a fashion similar to an excitation wave in excitable medium~\cite{adamatzky_delacycostello_shirakawa}. When plasmodium approaches domains occupied by tablets it avoids close contact with agar gel nearby the tablets. Thus the plasmodium's growth-front splits onto several wave-fragments (Fig.~\ref{nutrientsubstrate}). Depending on a physiological state of the plasmodium and contamination of its substrate by other moulds and bacteria the plasmodial wave-fragments either propagate as isolated fragments or merge later in a single wave-front. 
The plasmodium reacts to the tablets as if they were sources of repellents, e.g. 
inorganic salts~\cite{ueda_1976,adamatzky_physarum_salt,adamatzky_physarummachines}. We did 
not observe any attraction of plasmodium towards tablets in our experiments with nutrient agar gel.  

\begin{finding}
Herbal sleep-remedy tablets are treated by plasmodium of P. polycephalum as attractants when the plasmodium grows on a non-nutrient substrate.
\end{finding}

\begin{figure}[!tbp]
\centering
\subfigure[]{\includegraphics[width=0.49\textwidth]{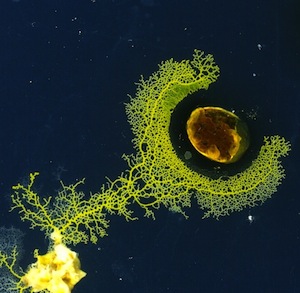}}
\subfigure[]{\includegraphics[width=0.49\textwidth]{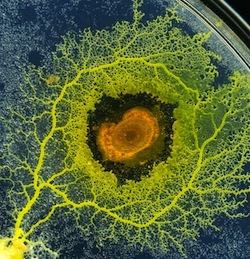}}
\subfigure[]{\includegraphics[width=0.49\textwidth]{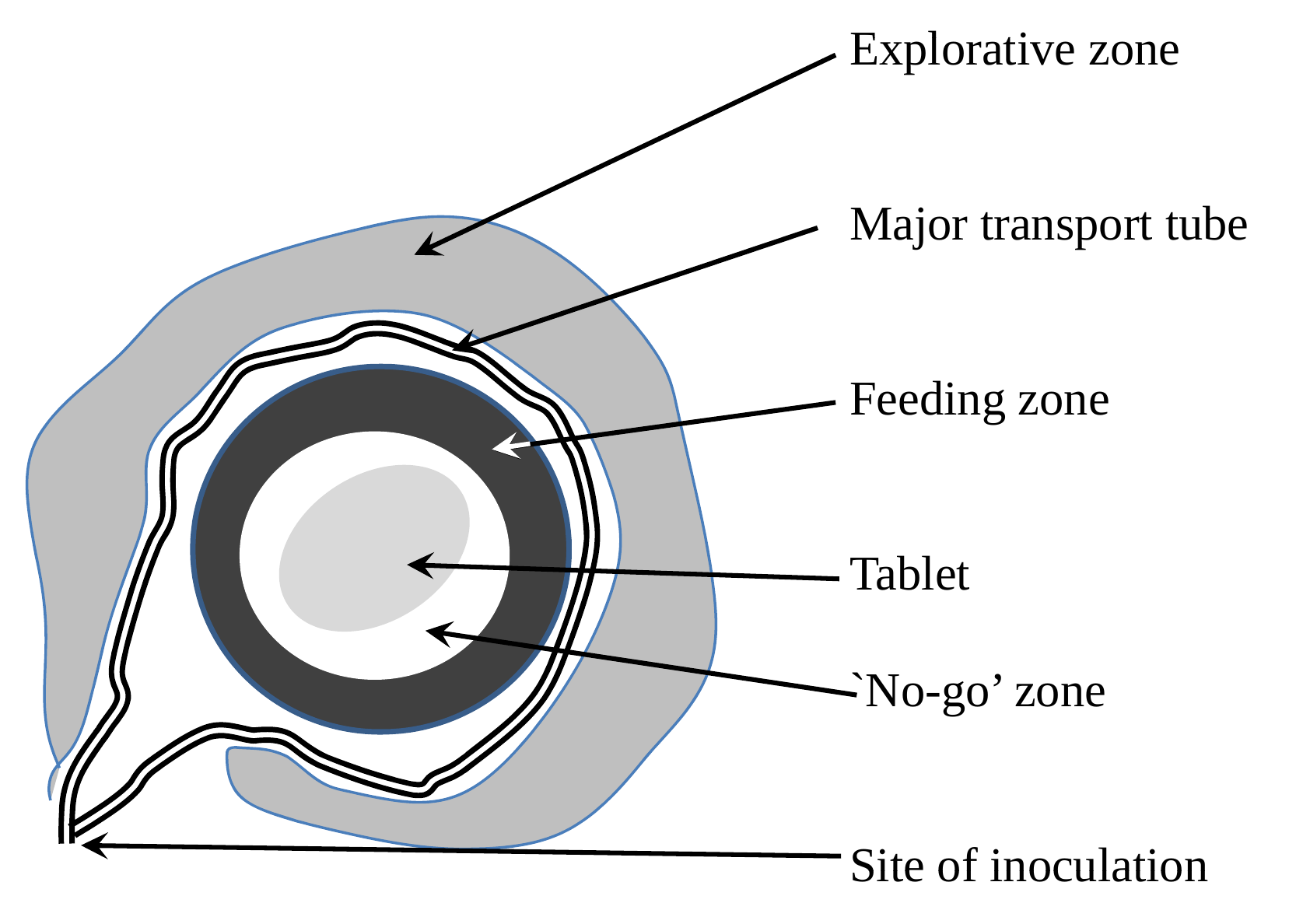}}
\subfigure[]{\includegraphics[width=0.49\textwidth]{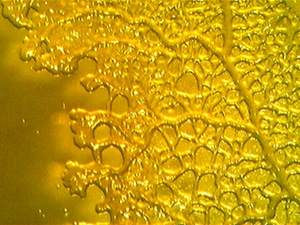}}
\caption{Interaction of the plasmodium growing on a non-nutrient substrate
with a single Kalms Sleep tablet. (ab)~Two stages of enveloping the tablet by plasmodium, scanned Petri dishes.
(c)~Scheme of principal zones in plasmodium-tablet complex. (d)~Photograph of the feeding zone, $\times$40. }
\label{singletablet}
\end{figure}

On a non-nutrient agar gel plasmodium propagates as localised excitation 
wave-fragments do in sub-excitable Belousov-Zhabotinsky medium~\cite{adamatzky_delacycostello_shirakawa}.
It starts growing in one directions, usually follows gradients of attractants and repellents, humidity and illumination. 
Each growing 'cone', or \emph{active zone}, the  analog of travelling localisation, has a certain degree of autonomy yet, coordinates its strategy of  space exploration with other parts of the plasmodium. In laboratory experiments we demonstrated that the plasmodium growing on a non-nutrient substrate is strongly attracted to the tablets but never comes into direct contact with the tablets or a substrate under the tablets. Further we are discussing experiments with the plasmodium growing on a non-nutrient substrate. 

Let us consider experiments with a single tablet (Fig.~\ref{singletablet}). Being inoculated at some distance from a tablet the plasmodium propagates directly towards the tablet. On approaching the tablet,  2-4~mm away form the tablet, the plasmodium generates two branches (Fig.~\ref{singletablet}a). One branch travels around the tablet clockwise another anti-clockwise. Eventually these two  propagating  fronts meet at the site opposite the site of the plasmodium's initial branching, and merge (Fig.~\ref{singletablet}b). A single tablet acts as a strong unequivocal attractive source in the plasmodium's foraging space. After enveloping the tablet with its body the plasmodium does not venture to any other domain of the substrate but stays virtually motionless. 

We can classify plasmodium-tablet complex on the major transport artery, feeding zone and exploratory zone (Fig.~\ref{singletablet}c). Major transport artery is a pronounced protoplasmic tube encircling the tablet. It separates 
plasmodium's body on two parts: feeding zone and exploratory zone. Feeding zone (Fig.~\ref{singletablet}d) consists of numerous branching processes staying at the edge of `no-go' zone, proximal to the tablet's domain not occupied by  plasmodium. 

\begin{figure}[!tbp]
\centering
\subfigure[12~h]{\includegraphics[width=0.32\textwidth]{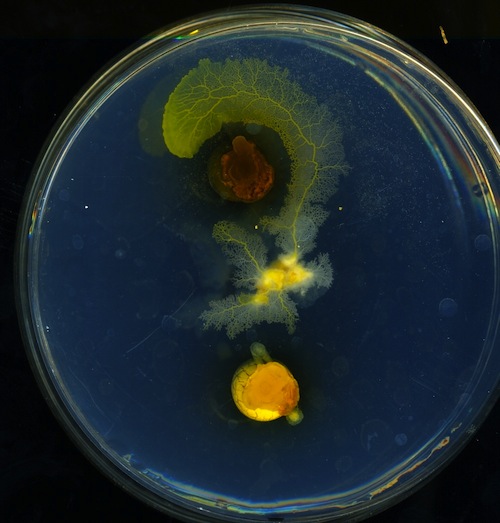}}
\subfigure[19~h]{\includegraphics[width=0.32\textwidth]{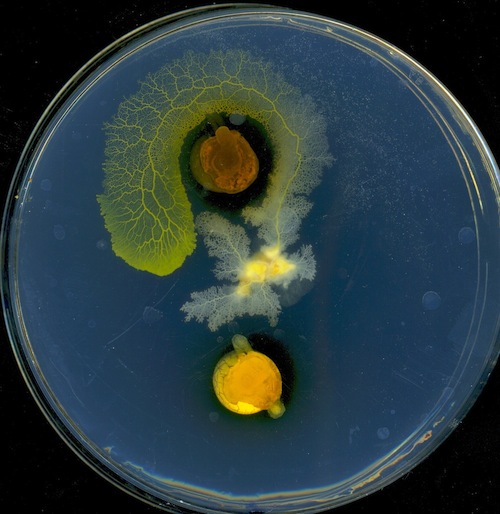}}
\subfigure[29~h]{\includegraphics[width=0.32\textwidth]{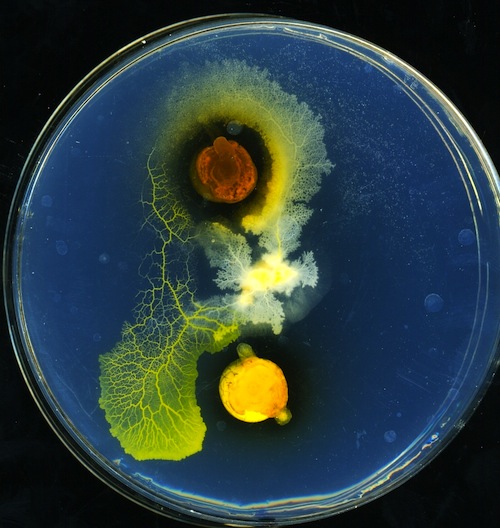}}
\subfigure[38~h]{\includegraphics[width=0.32\textwidth]{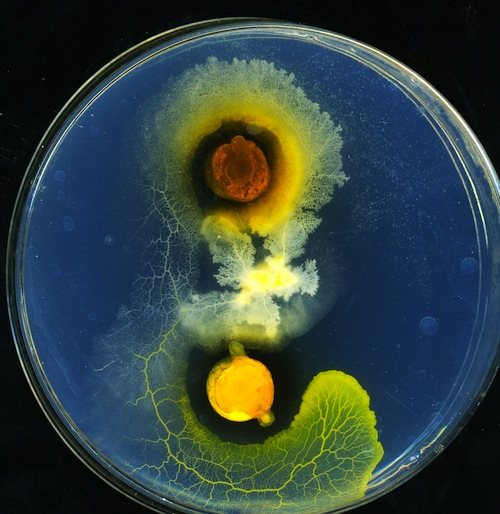}}
\subfigure[47~h]{\includegraphics[width=0.32\textwidth]{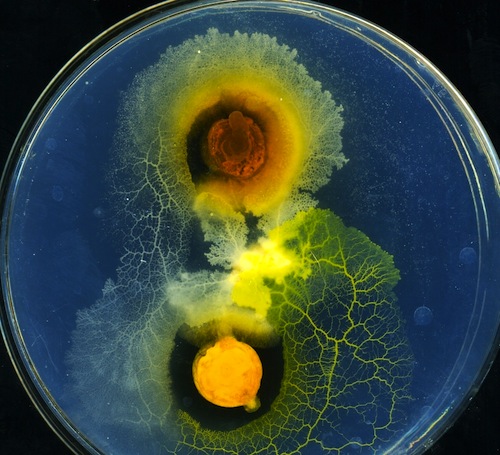}}
\subfigure[]{\includegraphics[width=0.32\textwidth]{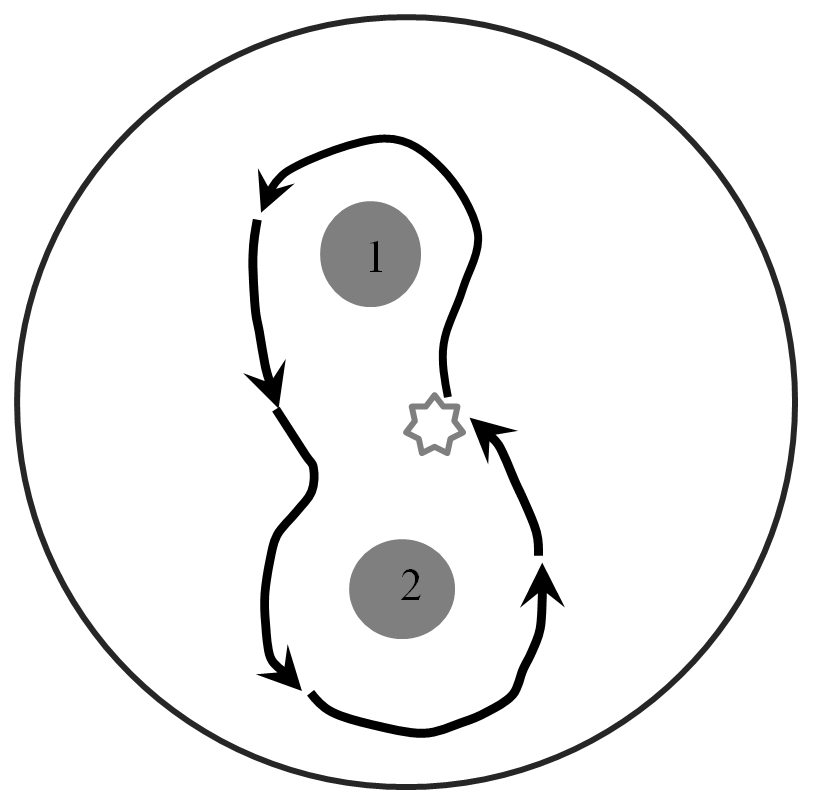}}
\caption{Plasmodium, on a non-nutrient substrate, navigates around two Kalms Sleep (north part of Petri dish) and Kalms Tablet tablets (south part of Petri dish). (a--e)~Snapshots of experimental Petri dish taken at various time intervals, indicated in hours from the moment of inoculation. (f)~Scheme of the plasmodium propagation: tablets are grey discs, inoculation site is the star and the trajectory of plasmodium's propagation is shown by arrows. }
\label{excellentrotation}
\end{figure}

\begin{finding}
A single herbal tablet acts as a fixed attractor for plasmodium of P. polycephalum while a sparse group of several
herbal tablets may act as limit cycle.
\end{finding}
 
 If two or more tablets  are present in a Petri dish and `no-go zones' of the tables do not intersect, the plasmodium travels between the tablets in a periodic fashion. Here we consider three most common situations. Plasmodium travelling along a limit cycle around two tablets is shown in Fig.~\ref{excellentrotation}. Kalms Sleep tablet is positioned at the north part of Petri dish and Kalms Tablets tablet at the south part. Plasmodium is inoculated in the centre of the Petri dish. In first 12~h after  inoculation the plasmodium propagates north and encounters Kalms Sleep tablet (Fig.~\ref{excellentrotation}a). The plasmodium makes anti-clockwise turn around the tablet (Fig.~\ref{excellentrotation}b) and travels south until it reaches  Kalms Tablets tablet (Fig.~\ref{excellentrotation}c). The plasmodium travels around the tablet anti-clockwise (Fig.~\ref{excellentrotation}d) and heads north towards Kalms Sleep tablet (Fig.~\ref{excellentrotation}e). If there was no accumulation of plasmodium's metabolites in agar gel the plasmodium would travel around the tablets non-stop (Fig.~\ref{excellentrotation}f).

\begin{figure}[!tbp]
\centering
\subfigure[13~h]{\includegraphics[width=0.45\textwidth]{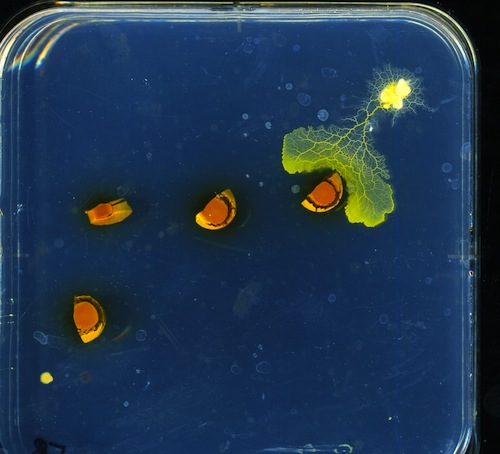}}
\subfigure[40~h]{\includegraphics[width=0.45\textwidth]{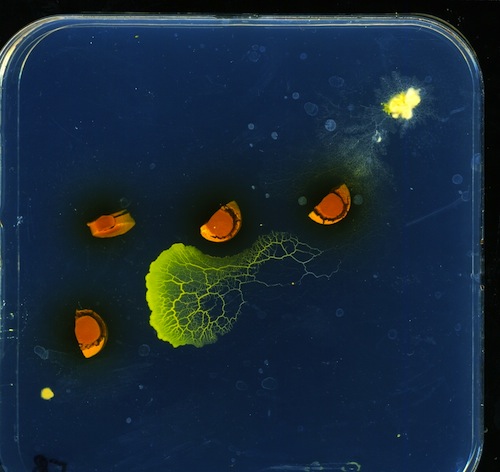}}
\subfigure[52~h]{\includegraphics[width=0.45\textwidth]{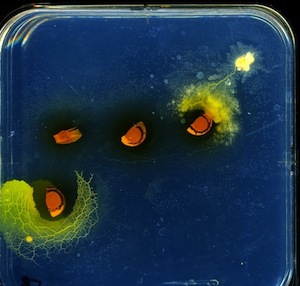}}
\subfigure[68~h]{\includegraphics[width=0.45\textwidth]{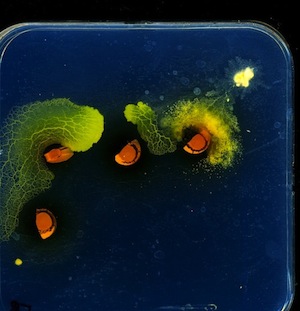}}
\subfigure[90~h]{\includegraphics[width=0.45\textwidth]{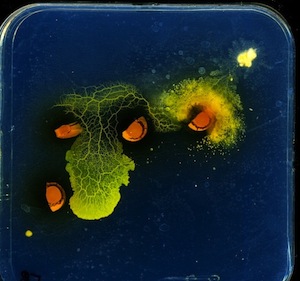}}
\subfigure[113~h]{\includegraphics[width=0.45\textwidth]{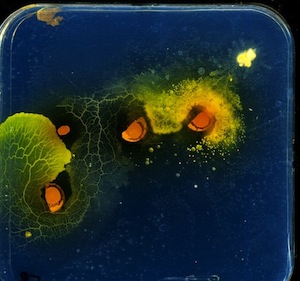}}
\caption{Formation of the `limit cycle' supported by streams of plasmodium wave-fronts, on a non-nutrient substrate.
(a--f)~Snapshots of plasmodia propagating in the configuration of Nytol tablets.}
\label{coolrotation}
\end{figure}

\begin{figure}[!tbp]
\centering
\subfigure[]{\includegraphics[width=0.45\textwidth]{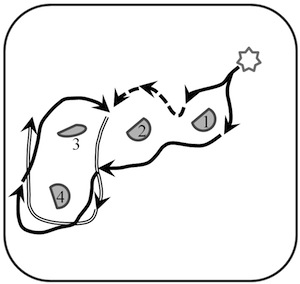}}
\subfigure[]{\includegraphics[width=0.45\textwidth]{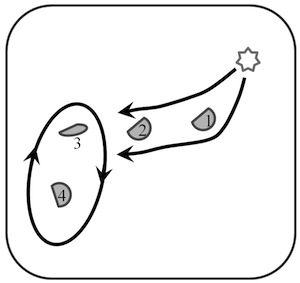}}
\caption{Schematic formation of the `limit cycle' supported by streams of plasmodium wave-fronts, on a non-nutrient substrate. (a)~Trajectories of plasmodia propagation, solid line indicates plasmodia propagation during first 60~h after inoculation, dotted line shows trajectory of plasmodium emerged and developed during period 68-90~h, and double line reflects propagation during 90-113~h. (b)~An idealised situation when active zones rotate around the tablets fed by 
streams of plasmodia. The schemes are derived from experimental data presented in Fig.~\ref{coolrotation}.}
\label{coolrotationscheme}
\end{figure} 

In example shown in Fig.~\ref{excellentrotation} plasmodium keeps its structural integrity during circumnavigation. Active zone remains connected to the rest of plasmodium's body with a network of protoplasmic tubes. Let us now consider experiment where plasmodium splits into several independent plasmodia, each one with its own active zone (Fig.~\ref{coolrotation}).  While describing plasmodia's behaviour we will refer to  Nytol tablets numbered from 1 to 4 as in scheme Fig.~\ref{coolrotation}gh. At the beginning of experiment an oat flake colonised by plasmodium is placed in north-east corner of a rectangular Petri dish. In 13~h after inoculation an active zone is formed. The active zone propagates towards tablet 1 and splits into two active  zones (Fig.~\ref{coolrotation}a). One active zone does not survive, another travels around tablet 2 and heads towards  tablet 4 (Fig.~\ref{coolrotation}b). The active zone reaches tablet 4 and starts moving clockwise around the tablet by 52nd~h   of experiment (Fig.~\ref{coolrotation}c). After passing tablet 4 along its western side the plasmodium reaches tablet 3 and   turns around the table 3 clockwise (Fig.~\ref{coolrotation}d). At roughly the same time the plasmodium still residing on the colonised oat flake emits new active zone which travels north of tablets 1 and 2 towards tablet 3 (Fig.~\ref{coolrotation}d). The active zone moving around tablet 3 and heading east-south-east collides with active zone travelling west around tablet 2. In the result of such collision two active zones merge and form a single active zone travelling south (Fig.~\ref{coolrotation}e). This active zone reaches tablet 4, turns around the table clockwise and travels north (Fig.~\ref{coolrotation}f). 
The scheme of propagation is shown in Fig.~\ref{coolrotationscheme}a. In an ideal situation, the system plasmodia-Nytol tablets would form a carousel of active zones rotating around tablets 3 and 4 and supported by streams of plasmodia periodically emitted from the site of inoculation (Fig.~\ref{coolrotationscheme}b).

\begin{figure}[!tbp]
\centering
\subfigure[12~h]{\includegraphics[width=0.45\textwidth]{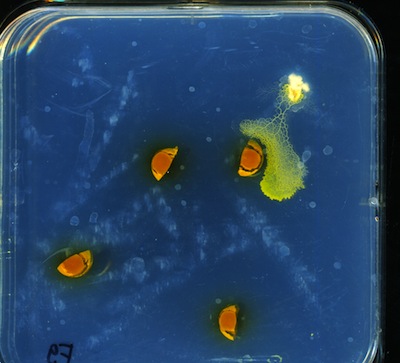}}
\subfigure[37~h]{\includegraphics[width=0.45\textwidth]{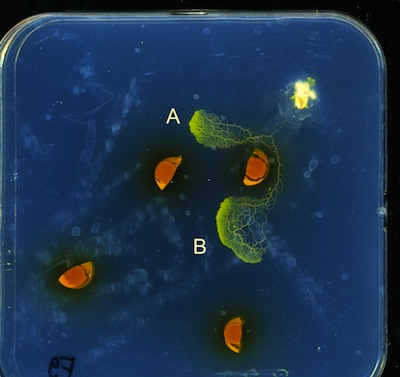}}
\subfigure[49~h]{\includegraphics[width=0.45\textwidth]{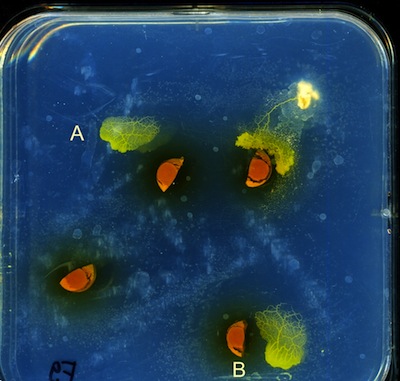}}
\subfigure[61~h]{\includegraphics[width=0.45\textwidth]{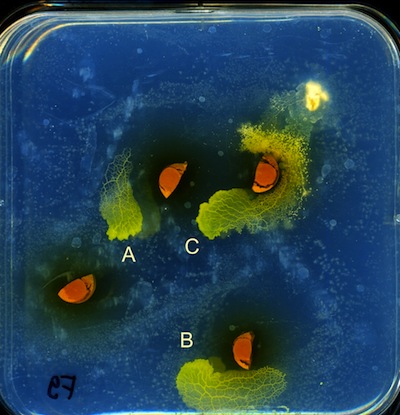}}
\subfigure[85~h]{\includegraphics[width=0.45\textwidth]{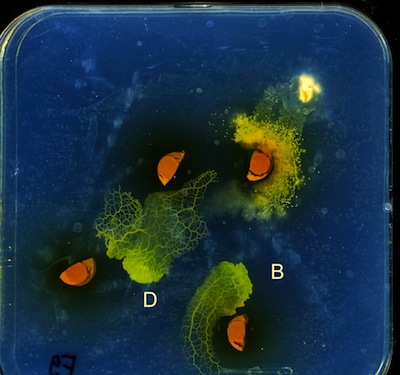}}
\subfigure[109~h]{\includegraphics[width=0.45\textwidth]{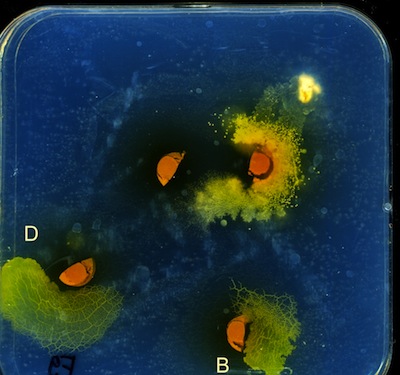}}
\caption{Complex navigation of plasmodia in the configuration of four Nytol tablets, on a non-nutrient substrate. 
(a--f)~Snapshots of experimental Petri dish.}
\label{coolfragments}
\end{figure}

\begin{figure}[!tbp]
\centering
\includegraphics[width=0.45\textwidth]{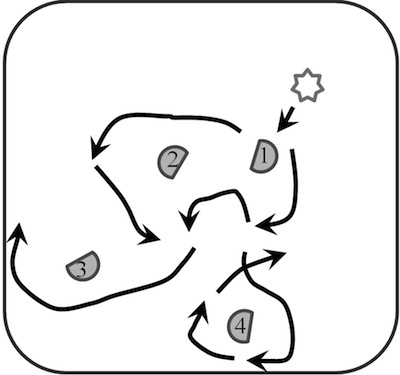}
\caption{Scheme of complex navigation of plasmodia in the configuration of four Nytol tablets, on a non-nutrient substrate, derived from experimental data in Fig.~\ref{coolfragments}.}
\label{coolfragmentsscheme}
\end{figure} 

Example of complex navigation of plasmodia around configuration of Nytol tablets is demonstrated in 
Fig.~\ref{coolfragments}). An oat flake colonised by plasmodium is placed in north-east corner of rectangular Petri dish. The plasmodium propagates towards tablet 1 (see labelling of tablets in Fig.~\ref{coolfragments}g) and splits into two active zones $A$ and $B$ (Fig.~\ref{coolfragments}ab). In 40-50~h after inoculation these active zones become disconnected and travel as independent plasmodia (Fig.~\ref{coolfragments}bc). Plasmodium $A$ travels north of tablet 2 and plasmodium $B$ heads towards tablet 4 (Fig.~\ref{coolfragments}c). At the same time additional active zone $C$ is emitted from the oat flake colonised by plasmodium and propagates towards tablet 1 (Fig.~\ref{coolfragments}d). By 60th hour from the beginning of experiment plasmodium $B$ makes almost complete clockwise rotation around tablet 4 while plasmodium $A$ makes anti-clockwise rotation around  tablet 2 (Fig.~\ref{coolfragments}d). Eventually plasmodia $A$ and $C$ collide, merge and propagate further as a single  plasmodium $D$ (Fig.~\ref{coolfragments}e). This plasmodium $D$ rotates clockwise around tablet 3 and plasmodium $B$ continues its clockwise rotation around tablet 4 (Fig.~\ref{coolfragments}f). See trajectories of active zones and plasmodia in Fig.~\ref{coolfragmentsscheme}.

\section{Numerical simulation}
\label{simulation}

A profile of plasmodium's active zone growing on a non-nutrient substrate is isomorphic to shapes 
of wave-fragments in sub-excitable media~\cite{adamatzky_delacycostello_shirakawa}. When active 
zone of \emph{P. polycephalum} spreads, two processes occur simultaneously --- advancing of 
the wave-shaped tip of the pseudopodium and formation of the trail of protoplasmic tubes. 
We simulate the chemo-tactic travelling of plasmodium using two-variable Oregonator 
equations~\cite{field_noyes_1974,tyson_fife}:

$$\frac{\partial u}{\partial t} = \frac{1}{\epsilon} (u - u^2 - (f v + \phi)\frac{u-q}{u+q}) + D_u \nabla^2 u$$
$$\frac{\partial v}{\partial t} = u - v .$$

The variable $u$ is abstracted as a local density of plasmodium's protoplasm and $v$ reflects local concentration of 
metabolites and nutrients. We integrate the system using Euler method with five-node Laplace operator, time step $\Delta t=5\cdot10^{-3}$ and grid point spacing $\Delta x = 0.25$, with the following parameters: $\phi=\phi_0 - \eta/2$, $A=0.0011109$, $\phi_0=0.0766$ for imitating plasmodium growth on a non-nutrient substrate and $\phi_0=0.066$ for nutrient-substrate, $\epsilon=0.03$, $f=1.4$, $q=0.022$.

Parameters $q$ and $f$ are inherited from the Oregonator model of Belousov-Zhabotinsky medium~\cite{field_noyes_1974,tyson_fife},
$\phi$ is proportional to local concentration of attractants and repellents. The parameter
 $\eta$ corresponds to a gradient of chemo-attractants emitted by data planar points. 
Let $\bf P$ be a set of attraction sites $\mathbf P$ and $x$ be a site of a simulated medium then  
$ \eta_x = 2 \cdot 10^{-2} - \min_{p \in \bf P}  \{ d(x,p): \gamma(p)=$ {\sc True}$ \} \cdot b^{-1} $
where $3.1 \cdot 10^{2} \leq b \leq 4.9 \cdot 10^{2}$ and $d(x,p)$ (for the simulated medium $400\times400$ sites) 
is an Euclidean distance between sites $x$ and $p$. 

We imitate narrow feeding zones around tablets (Fig.~\ref{singletablet}ab) as follows. Let $x$ be a centre
of a tablet radius $r>0$, we assume $\phi_0=0.065$ in any site $y$ such that $ r \geq d(y,x) \leq r+w$, 
where $w'$ is a the feeding zone's width.

The medium is perturbed by an initial excitation, where a $11\times11$ sites are assigned $u=1.0$ each. The perturbation generates a propagation wave-fragment travelling along gradient $\eta$.  Repellents emitted by sites 
of $\mathbf P$ show very limited diffusion. Therefore repellents can be regarded as impassable obstacles.

To imitate formation of the protoplasmic tubes we store values of $u$ in matrix $\bf L$, which is processed at the end of simulation. For any site $x$ and time step $t$ if $u_x>0.1$ and $L_x=0$ then $L_x=1$. The matrix $\bf L$ represents time lapse superposition of propagating wave-fronts. The simulation is considered completed when propagating pattern envelops $\mathbf P$ and halts any further motion.  At the end of simulation we repeatedly apply the erosion operation~\cite{adamatzky_physarummachines},  which represents  a stretch-activation effect~\cite{kamiya_1959} necessary for formation of plasmodium tubes, to $\bf L$. The resultant protoplasmic network  provides a good phenomenological match for networks recorded in laboratory experiments. 

\begin{figure}[!tbp]
\centering
\subfigure[]{\includegraphics[width=0.44\textwidth]{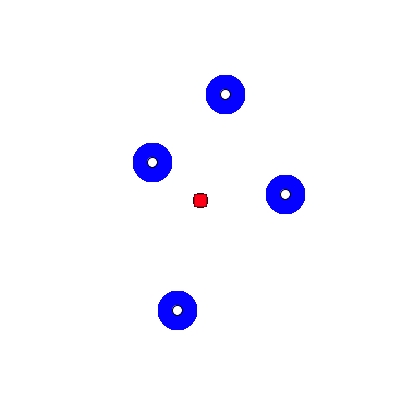}}
\subfigure[]{\includegraphics[width=0.44\textwidth]{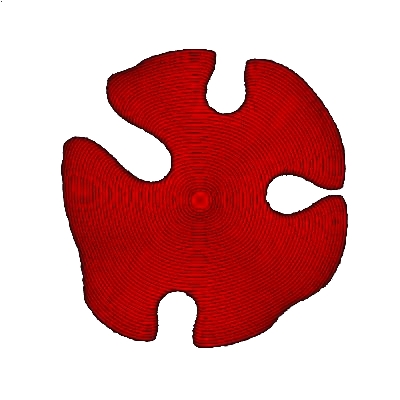}}
\caption{Simulation of plasmodium propagation on a nutrient substrate with four tablets.
(a)~Initial configuration, tablets are shown as blue/dark grey washers and initial position of plasmodium is a 
red/light grey ellipse. (b)~Time-lapse snapshots of propagating plasmodium's front.  }
\label{nutrientrichmodel}
\end{figure} 

Our experimental finding on behaviour of plasmodium on a nutrient substrate (Fig.~\ref{nutrientsubstrate}) 
is fully verified in Oregonator computer model. The tablets are treated as obstacles and no attraction of 
plasmodium wave-fronts to the tablets is observed in computer simulations (Fig.~\ref{nutrientrichmodel}). 

\begin{figure}[!tbp]
\centering
\subfigure[]{\includegraphics[width=0.4\textwidth]{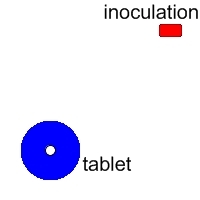}}
\subfigure[]{\includegraphics[width=0.4\textwidth]{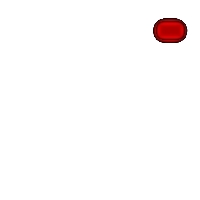}}
\subfigure[]{\includegraphics[width=0.4\textwidth]{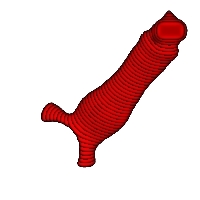}}
\subfigure[]{\includegraphics[width=0.4\textwidth]{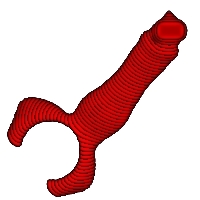}}
\subfigure[]{\includegraphics[width=0.4\textwidth]{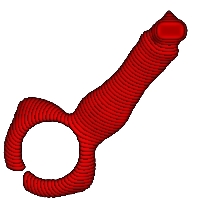}}
\subfigure[]{\includegraphics[width=0.4\textwidth]{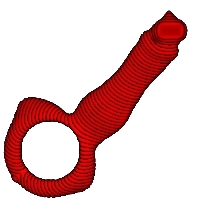}}
\caption{Simulation of plasmodium enveloping a single tablet on a non-nutrient substrate. (a)~Initial setup. 
(b--f)~Time-lapse images of propagating active zones. Zone of repellents is shown by blue/grey in (a).}
\label{onetabletsimulation}
\end{figure}

\begin{figure}[!tbp]
\centering
\subfigure[]{\includegraphics[width=0.46\textwidth]{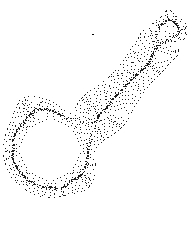}}
\subfigure[]{\includegraphics[width=0.46\textwidth]{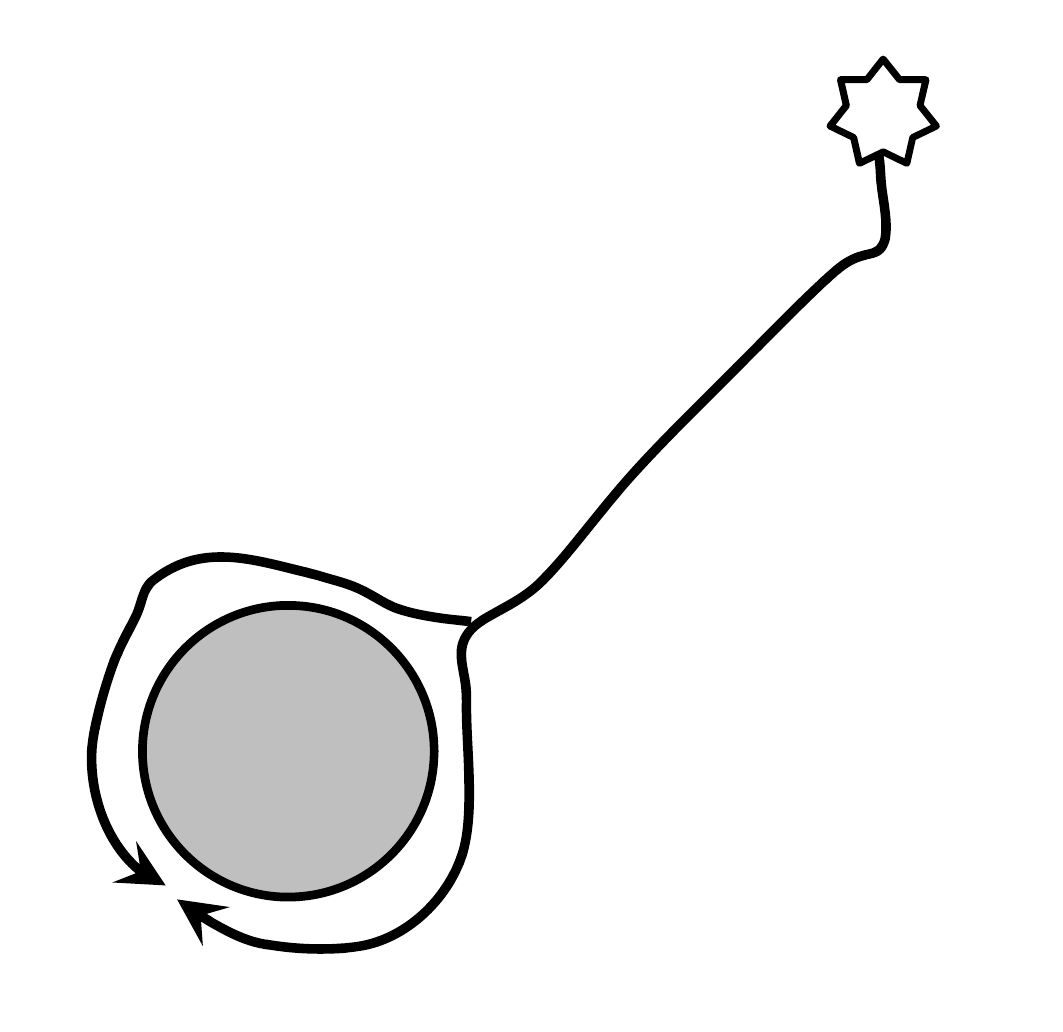}}
\caption{Simulation of plasmodium enveloping a single tablet on a non-nutrient substrate. 
(a)~Major protoplasmic network extracted from time-lapse images. 
(b)~Scheme of interaction of active growth zones. 
Inoculation site is shown by star in (b). 
Trajectories of plasmodium active zones' propagations are shown by arrows.}
\label{onetabletsimulationtubescheme}
\end{figure} 

Simulation of plasmodium enveloping a single table is shown in Fig.~\ref{onetabletsimulation}. We generate
a source of attractants at the south-west corner of a simulation space and perturb the medium in a $15 \times 10$ sites domain in  the north-east corner of the simulation space (Fig.~\ref{onetabletsimulation}a). Active zones, visible 
as fronts of wave-fragments in Fig.~\ref{onetabletsimulation}b--f travel along attractant gradients towards the tablet. 
On approaching the tablet the plasmodium's active zones encounter repellents and therefore do not propagate close to the tablets but  remain in the feeding zone (Fig.~\ref{onetabletsimulation}cde) and encapsulate the repellent domain (Fig.~\ref{onetabletsimulation}f). When active zone growing clockwise collide with active zone growing anti-clockwise the zones merge (Fig.~\ref{onetabletsimulation}f). 

Network of major protoplasmic tubes extracted from lapse-time images (Fig.~\ref{onetabletsimulation}) is shown in Fig.~\ref{onetabletsimulationtubescheme}a. The network satisfactory approximates the protoplasmic network recorded in laboratory experiments (Fig.~\ref{singletablet}ab). Summary scheme of plasmodium's propagation towards and enveloping of a simulated tablet is shown in  Fig.~\ref{onetabletsimulationtubescheme}b.

\begin{figure}[!tbp]
\centering
\subfigure[]{\includegraphics[width=0.44\textwidth]{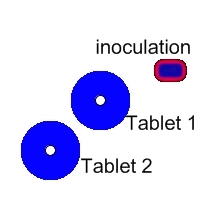}}
\subfigure[]{\includegraphics[width=0.44\textwidth]{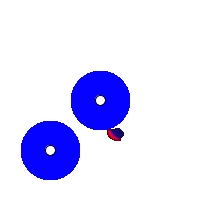}}
\subfigure[]{\includegraphics[width=0.44\textwidth]{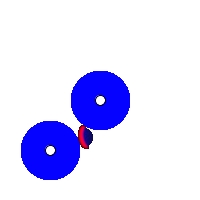}}\
\subfigure[]{\includegraphics[width=0.44\textwidth]{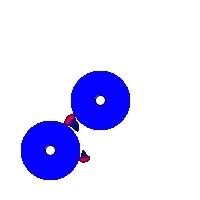}}
\subfigure[]{\includegraphics[width=0.44\textwidth]{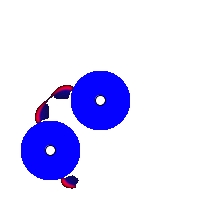}}
\subfigure[]{\includegraphics[width=0.44\textwidth]{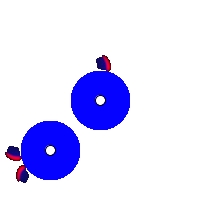}}\\
\caption{Simulation of the plasmodium orbiting around two tablets on a non-nutrient substrate.
(a)~Initial setup. 
(b)--(f)~Snapshots of wave-fragments, corresponding to tips of active zones.
(g)~Time-lapse image of propagating active zones.
}
\label{twotabletsimulation}
\end{figure} 

Simulated periodic motion of plasmodium's active zone around tablets is shown in Fig.~\ref{twotabletsimulation}. The simulation results are in agreement with experimentally observed motion patterns (Figs.~\ref{excellentrotation}, 
\ref{coolrotation} and \ref{coolfragments}). We arrange two simulated tablets as shown in Fig.~\ref{twotabletsimulation}a, and 
offset initial perturbation domain (analog of plasmodium's inoculation site) some distance away from 
a line passing through the tablets' centres. The tablets are arranged such that there is a space between the tablets exceeding double width of the feeding zones.

When plasmodium propagates close to tablet 1 the feeding zone splits in two parts, one part travels anti-clockwise another clockwise. Due to initial offset the zone travelling anti-clockwise annihilates and only active zone travelling clockwise continues its journey (Fig.~\ref{twotabletsimulation}b). When the active 
zone, travelling around tablet 1, approaches tablet 2  (Fig.~\ref{twotabletsimulation}c), it branches to a second active zone, which starts travelling clockwise around tablet 2 (Fig.~\ref{twotabletsimulation}d). The active zone travelling clockwise around tablet 1 then  finds itself in a close proximity of both tablets. Therefore the active zone splits into two propagating patterns: one active zone continues travelling clockwise around tablet 1 while second travels anti-clockwise around tablet 2 (Fig.~\ref{twotabletsimulation}e).

Therefore at some stage we observe three active zones (wave-fragments) circumnavigating the tablets: one active zone running clockwise around tablet 1, and two zones running anti- and clockwise around tablet 2 (Fig.~\ref{twotabletsimulation}ef). The active zones travelling around tablet 2 collide head-on (Fig.~\ref{twotabletsimulation}ef) and merge. The active zone travelling around tablet 1 continues its travel and the cycle of splitting/branching and annihilation is repeated from the step shown in Fig.~\ref{twotabletsimulation}b.

\begin{figure}[!tbp]
\centering
\subfigure[]{\includegraphics[width=0.44\textwidth]{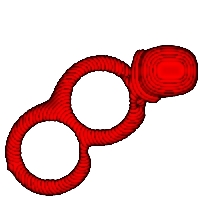}}
\subfigure[]{\includegraphics[width=0.44\textwidth]{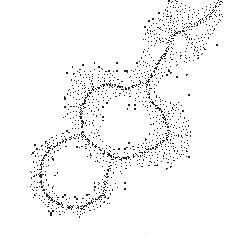}}
\subfigure[]{\includegraphics[width=0.44\textwidth]{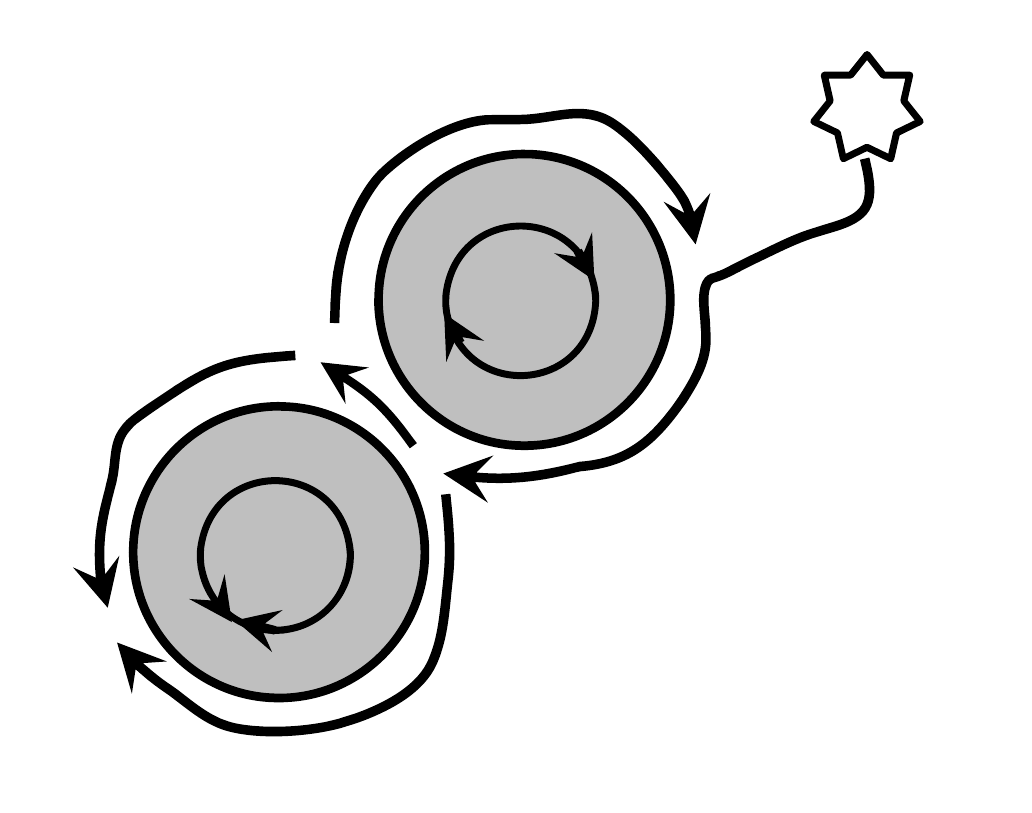}}
\caption{Simulation of the plasmodium orbiting around two tablets on a non-nutrient substrate.
(a)~Time-lapse image of propagating active zones.
(b)~Major protoplasmic network extracted from time-lapse images.
(c)~Scheme of periodic propagation of active zones.
}
\label{twotabletsimulationB}
\end{figure}

The static image of active zones' trajectories (Fig.~\ref{twotabletsimulationB}a) and the reconstructed network of protoplasmic tubes (Fig.~\ref{twotabletsimulationB}b) are similar to plasmodium shapes observed in laboratory experiments. See also a scheme of periodic movement in Fig.~\ref{twotabletsimulationB}c.  In laboratory experiments periodic movements never last more than 3-5 repetitions. Usually at some stage resources are depleted, or a substrate is contaminated with metabolites, and the plasmodium migrates to other parts of experimental arena.

\section{Possible applications of herbal tablets in computing with slime moulds}
\label{discussion}

In laboratory experiments and computer simulation we demonstrated that herbal tables may be in an untraditional ways of controlling spatial development of slime mould \emph{Physarum polycephalum}.  The slime mould is attracted to a single tablet, envelops a group of tightly packed tablets and circles between loosely positioned tablets. Possible active substances which could be responsible for such an unusual interaction between slime mould and herbal tablets are analysed in ~\cite{adamatzky_herbal}.  The herbal tablets is a cheap, user-friendly, and efficient alternative to existing techniques of controlling slime moulds with illumination-~\cite{nakagaki_yamada_1999}, thermo- \cite{tso_mansour_1975,matsumoto_1980}, and salt-based repellents~\cite{adamatzky_physarum_salt}, and carbohydrate-based attractants~\cite{carlile_1970,knowles_carlile_1978,dussutour_2010}. In present section we discuss two possible applications of sedative tables in slime mould based computing: approximation of planar shapes and implementation of 
logical gates. 

\subsection{Planar shapes}

\begin{figure}
\centering
\subfigure[]{\includegraphics[width=0.45\textwidth]{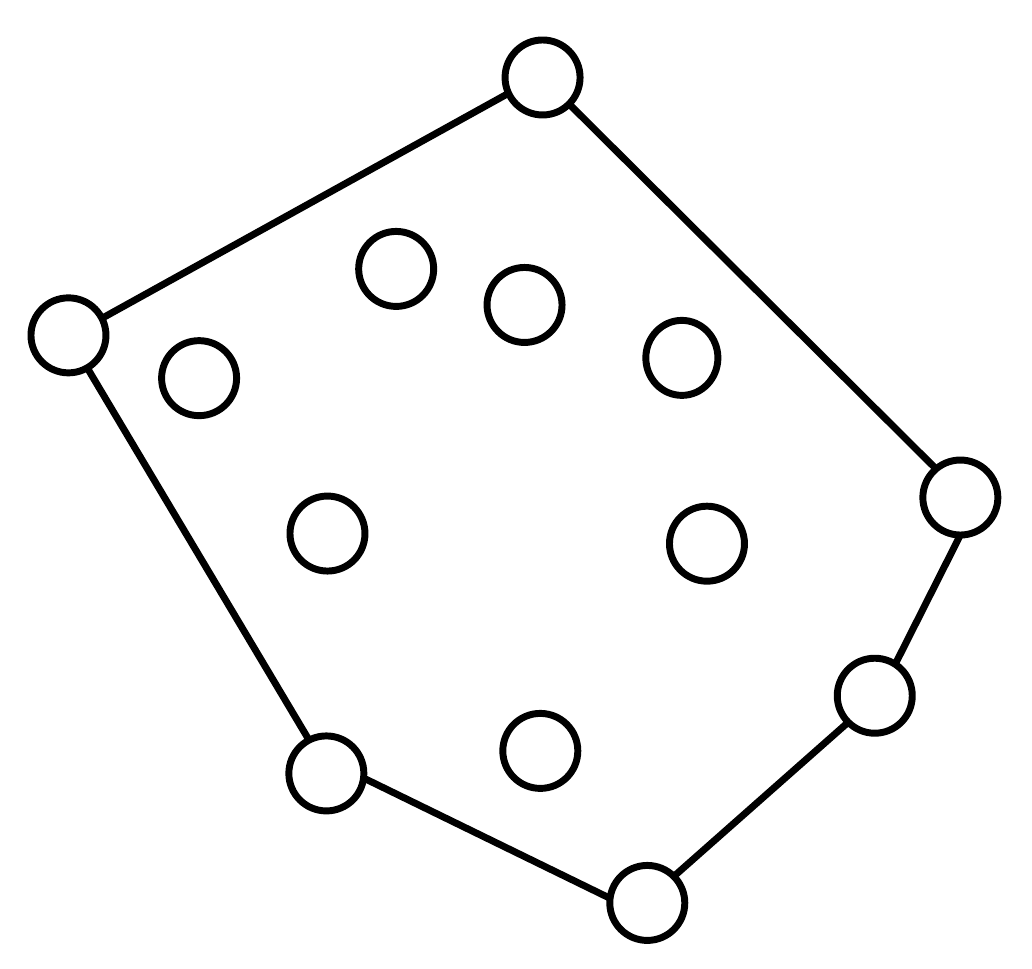}} 
\subfigure[]{\includegraphics[width=0.45\textwidth]{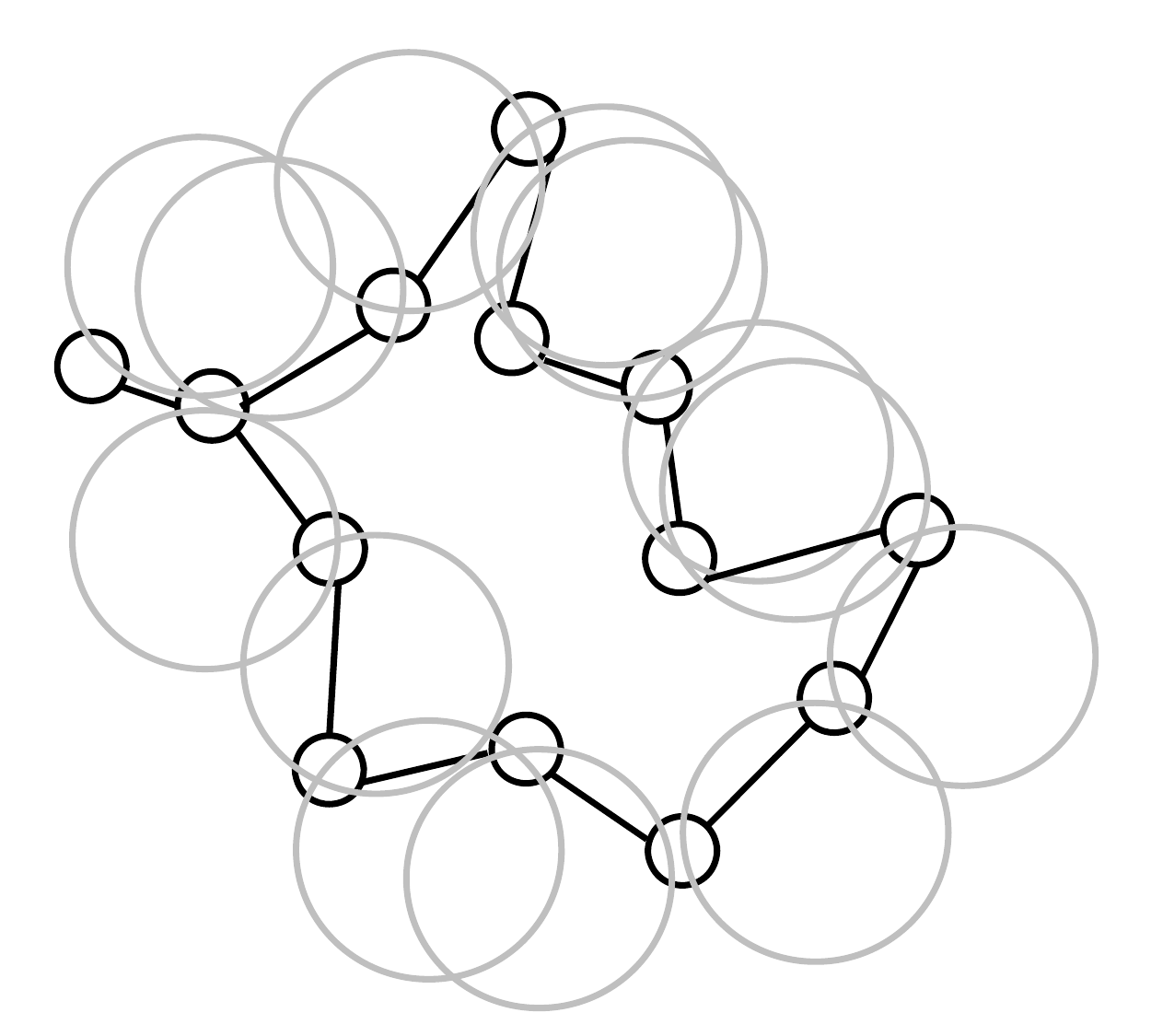}} 
\caption{Planar hulls. (a)~Example of convex hull. (b)~Example of concave hull. Points of set $\mathbf P$ are empty discs.
}
\label{basicshulls}
\end{figure}

A convex hull  of a finite set  $\mathbf P$  of planar points is the smallest convex 
polygon that contains all points  of $\mathbf P$ (Fig.~\ref{basicshulls}a). $\alpha$-hull of $\mathbf P$ 
is an intersection of the complement of all closed discs of radius $1/\alpha$ that includes no points 
of $\mathbf P$~\cite{edelsbrunner_1983,edelsbunner_1994}. $\alpha$-shape is a convex hull  
when $\alpha \rightarrow \infty$. With decrease of $\alpha$ the shapes may shrink, develop holes 
and become disconnected, the shapes collapse to $\mathbf P$  when $\alpha \rightarrow 0$. 
A concave hull is  non-convex polygon representing area occupied by $\mathbf P$. A concave hull is a connected
$\alpha$-shape without holes (Fig.~\ref{basicshulls}b). We want slime mould to solve the following problem. Given 
planar set $\mathbf P$ represented by physical objects plasmodium of \emph{P. polycephalum} must represent
concave hull of $\mathbf P$ by its largest protoplasmic tube.

\begin{figure}
\centering
\subfigure[]{\includegraphics[width=0.45\textwidth]{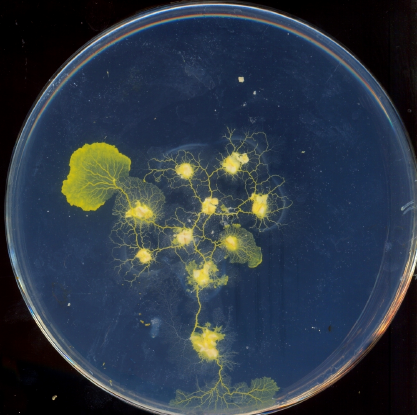}}         
\subfigure[]{\includegraphics[width=0.45\textwidth]{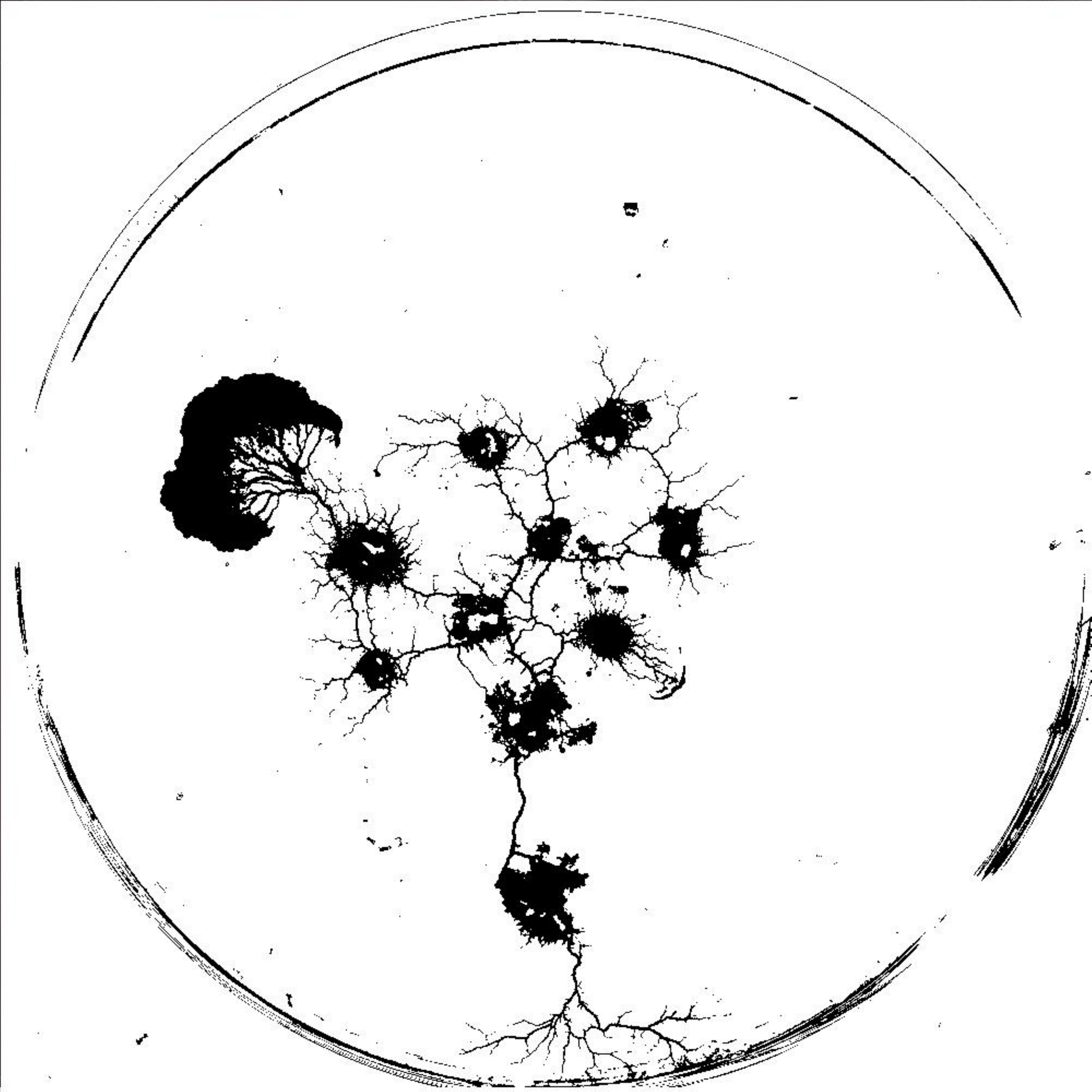}}   
\subfigure[]{\includegraphics[width=0.45\textwidth]{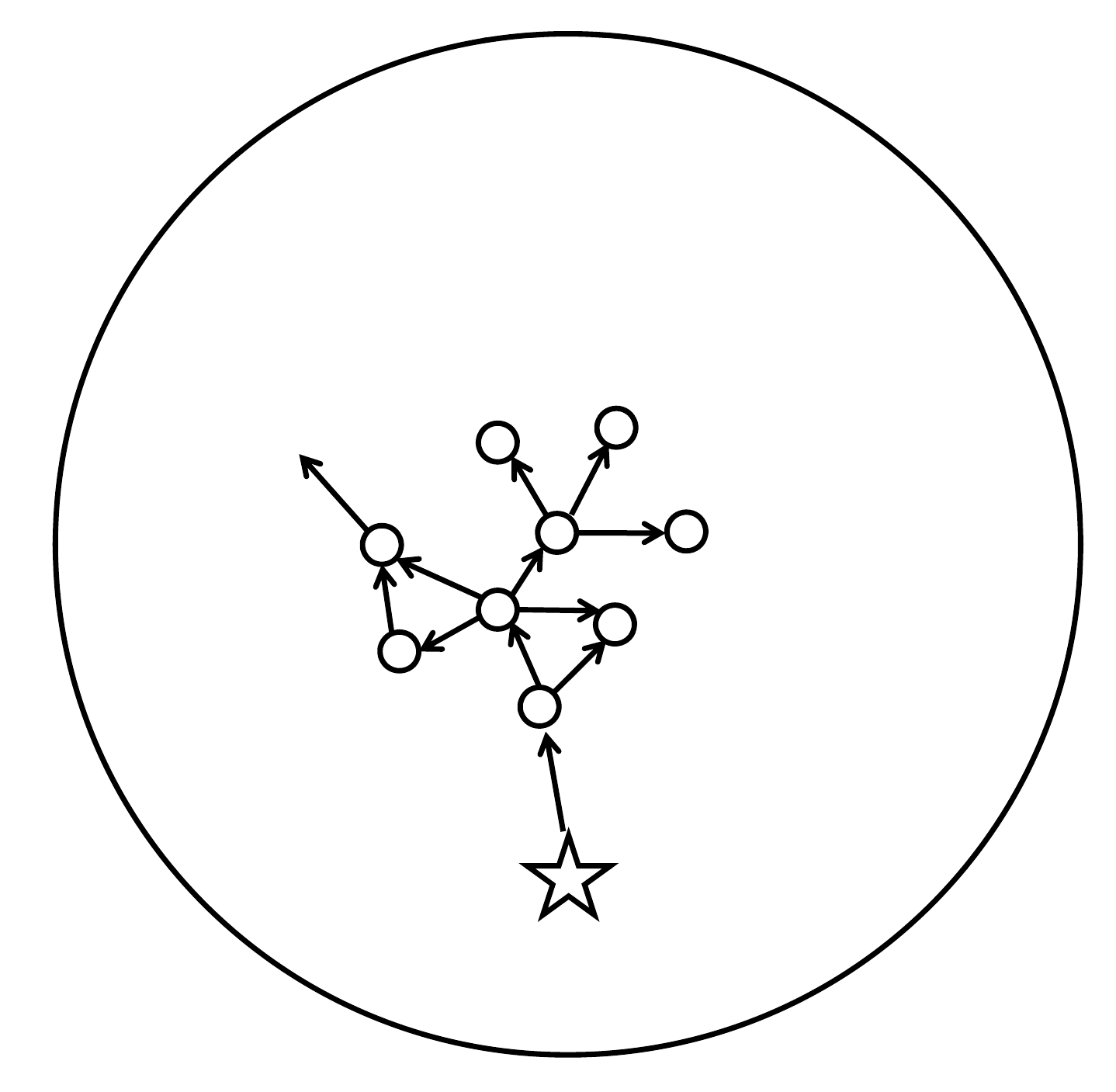}} 
\caption{Slime mould spans oat flakes.
 (ab)~Spanning oat flakes, representing $\mathbf P$, by a network of protoplasmic tubes:
 scanned image (a) and its binarisation (b).  (c)~Scheme of the plasmodium propagation, circles are points, or oat flakes, of $\mathbf P$ and  star marks site of inoculation, arrows are protoplasmic tubes.}
\label{basicsflakes}
\end{figure}

The fist algorithm of convex hull construction~\cite{jarvis_1973} was based on cognitive tactic techniques 
we use in our everyday's life. We select a starting point which is extremal point of $\mathbf P$. We pull 
a rope (anti-)clockwise to other extremal point. We continue until the set $\mathbf P$ is wrapped 
completely. The computation stops when we reach the starting point. Let we represent data points $\mathbf P$ by sources of attractants only, e.g. by oat flakes (Fig.~\ref{basicsflakes}a). We place a piece of plasmodium at some distance away from the set of points (Fig.~\ref{basicsflakes}a). The plasmodium propagates towards set $\mathbf P$, colonises oat flakes (Fig.~\ref{basicsflakes}ab) and spans them with a network of protoplasmic tubes (Fig.~\ref{basicsflakes}c). No hull is constructed. When all data points $\mathbf P$ 
are colonised and spanned by protoplasmic network the plasmodium ventures to explore the space around $\mathbf P$ and thus loses all chances of constructing any shapes of $\mathbf P$.

\begin{proposition}
P. polycephalum does not compute concave or convex hull of a set represented by attracting sources. 
\end{proposition}

\begin{figure}
\centering
\includegraphics[width=0.45\textwidth]{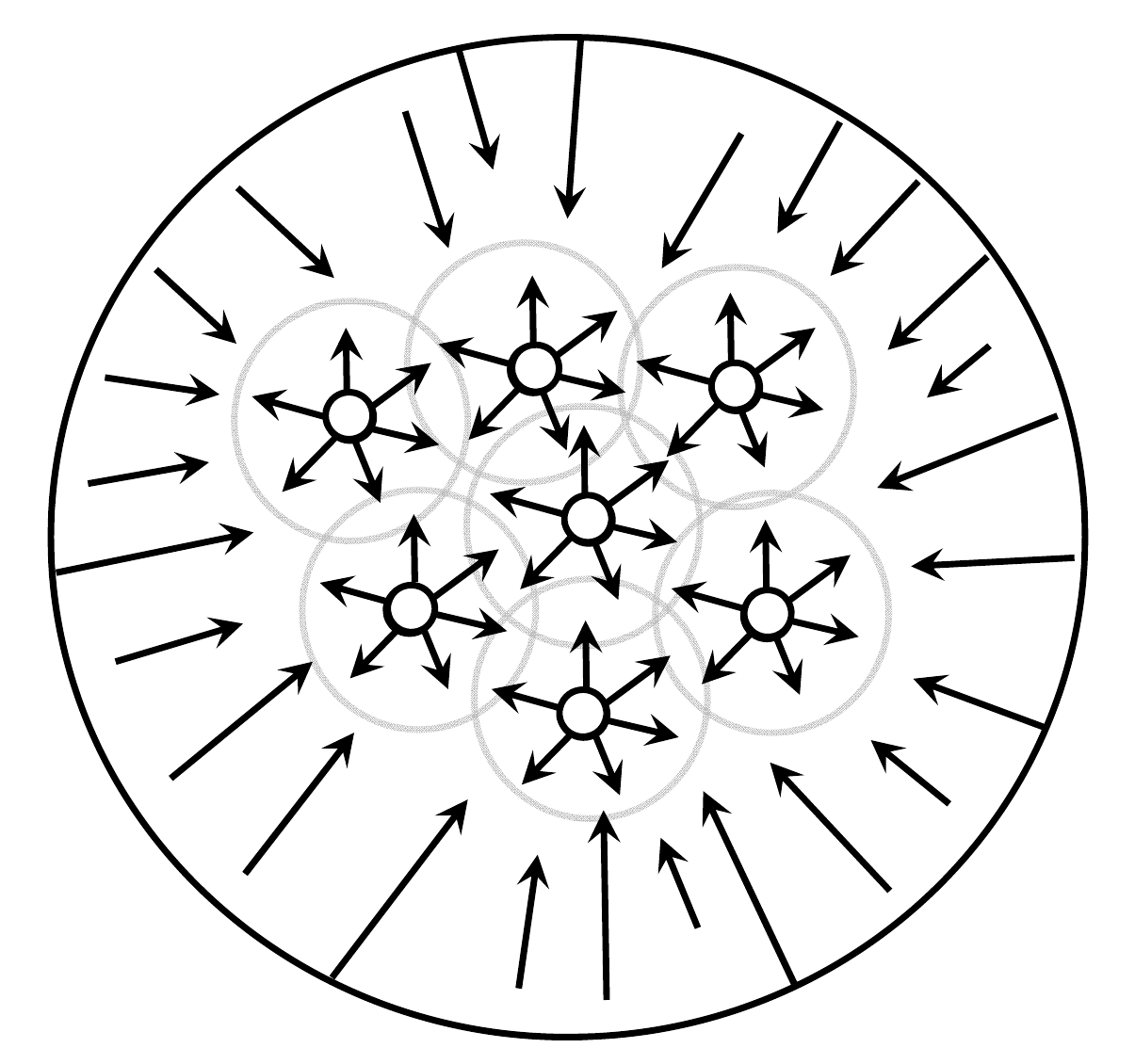}
\caption{Proposed distribution of attracting and repelling gradients which may 
force plasmodium to approximate a concave hull (arrows aiming towards set of discs are attractive forces, arrows originating in data points, discs, are repelling forces).}
\label{basicsscheme}
\end{figure}

Using repellents only, e.g. as a set of obstacles between attractants and inoculation site, will do no good because plasmodium will just pass around the repellents and leave (see details in \cite{adamatzky_physarummachines}, controlling Physarum with salt). The only solution would be to employ attractants to `pull' plasmodium towards planar set $\mathbf P$ and to use repellents to prevent plasmodium from spanning the points of $\mathbf P$ (Fig.~\ref{basicsscheme}). Strength of repellents should be proportional to $\alpha$ and thus will determine exact shape of the constructed hull. This corresponds to original definition~\cite{edelsbrunner_1983} that
$\alpha$-hull of $\mathbf P$ s the intersection of all closed discs with radius $l/\alpha$ that contain all the points
of $\mathbf P$.

\begin{proposition}
Plasmodium of P. polycephalum approximates connected $\alpha$-hull without holes 
of a finite planar set, which points are represented by sources of long-distance 
attractants and short-distance repellents. 
\end{proposition}

When presented with a half-pill of the Kalms Tablets/Sleep the plasmodium propagates towards the pill and forms, with its protoplasmic tubes, a circular enclosure around the pill. Such a unique behaviour of plasmodium in presence of Kalms Tablets/Sleep indicates that a plasmodium could implement Jarvis's Gift Wrapping algorithm~\cite{jarvis_1973}, adapted to concave hulls, if points of $\mathbf P$ are represented by the pills.

\begin{figure}
\centering
\subfigure[$t=12$~h]{\includegraphics[width=0.45\textwidth]{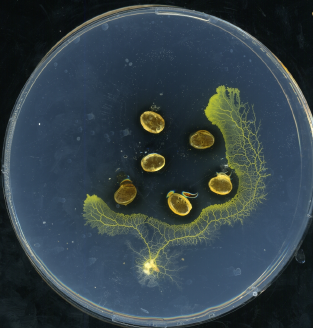}} 
\subfigure[$t=12$~h]{\includegraphics[width=0.45\textwidth]{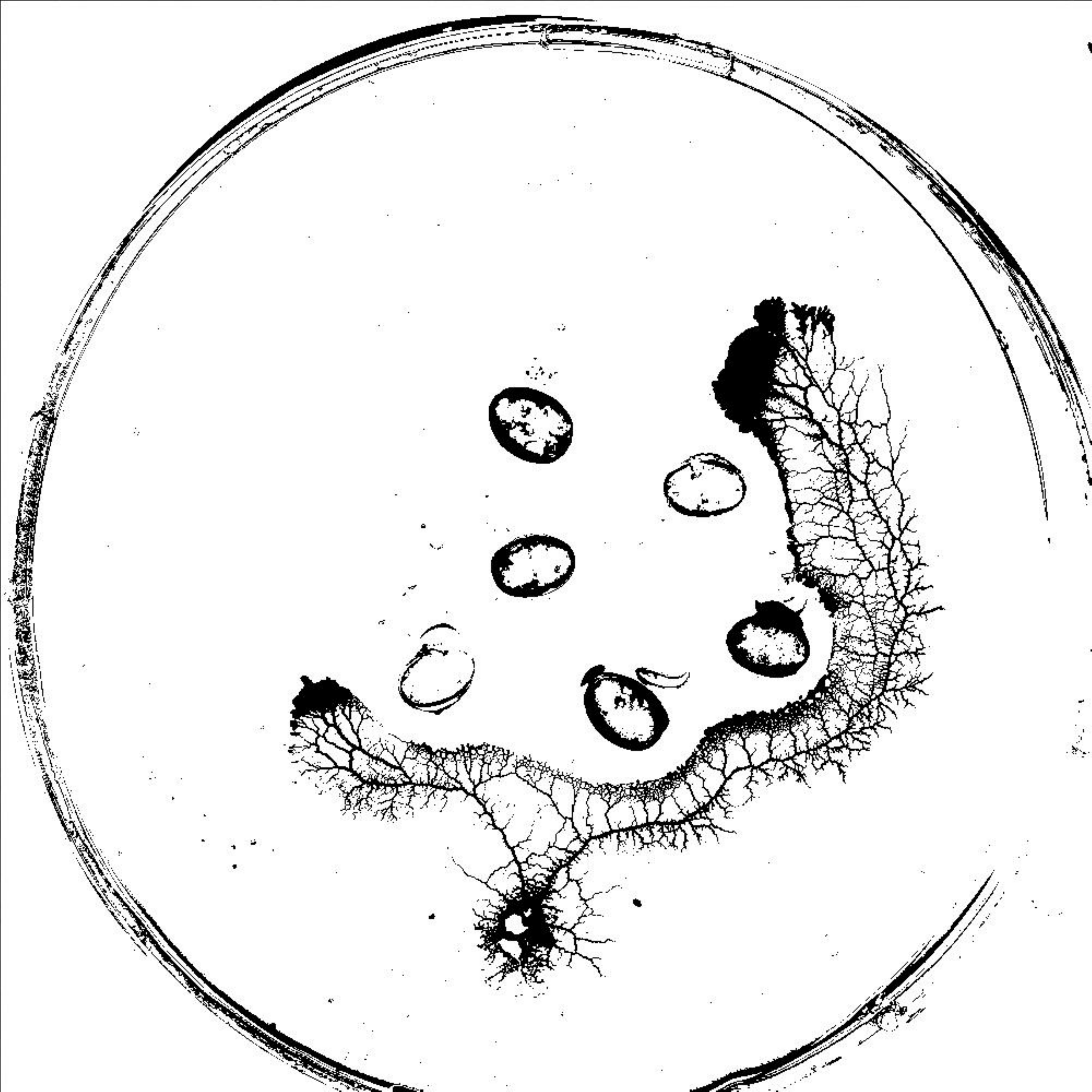}}
\subfigure[$t=24$~h]{\includegraphics[width=0.45\textwidth]{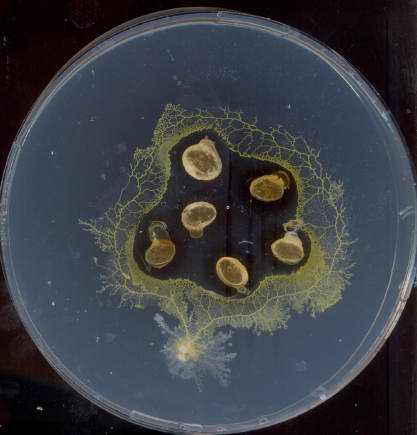}}
\subfigure[$t=24$~h]{\includegraphics[width=0.45\textwidth]{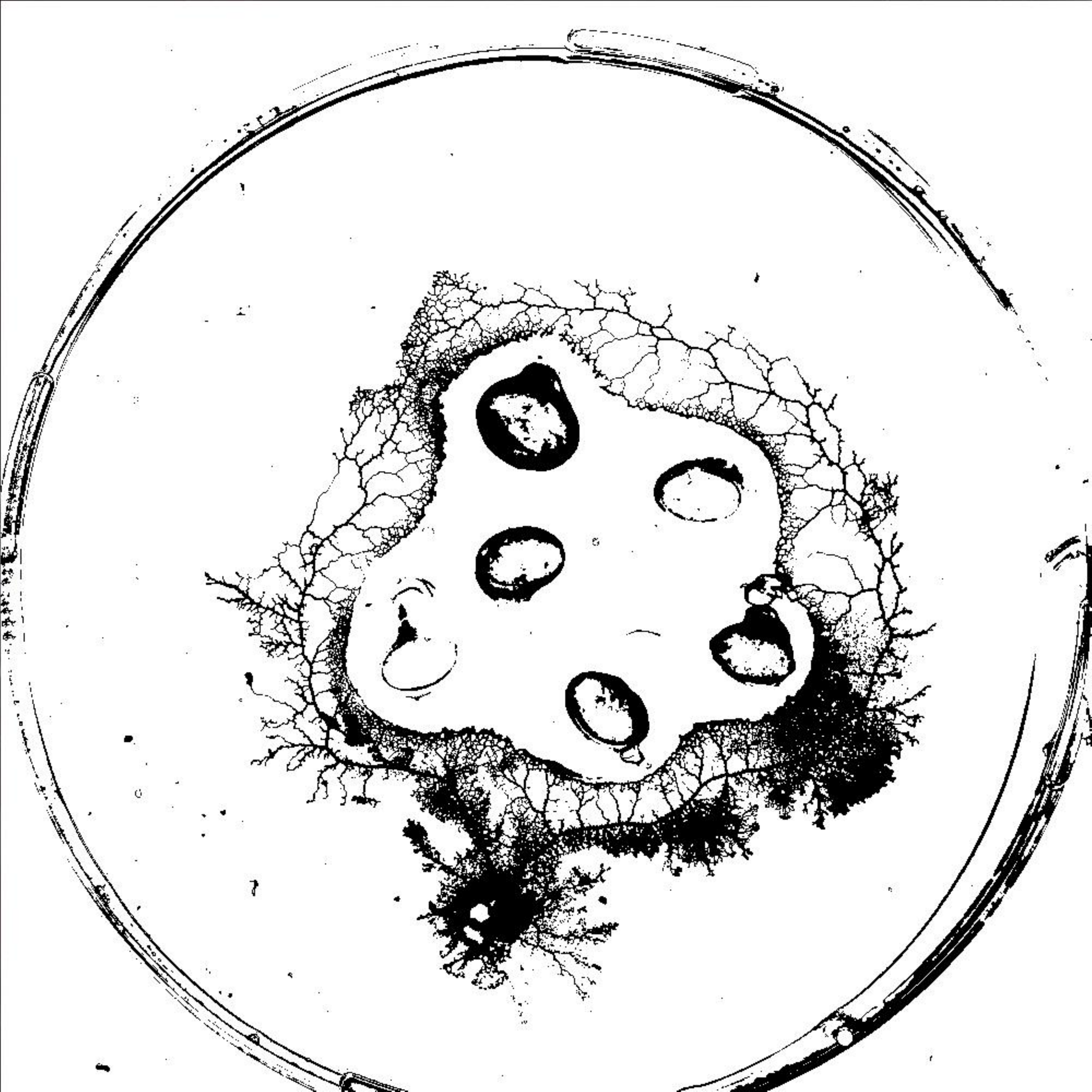}}
\caption{Computation of concave hulls of point set $\mathbf P$. Photographs~(ac) and 
their binarised images~(bd) taken 12~h~(ab) and 24~h~(cd) hours after 
plasmodium's inoculation. 
}
\label{example1}
\end{figure}

\begin{figure}
\centering
\includegraphics[width=0.45\textwidth]{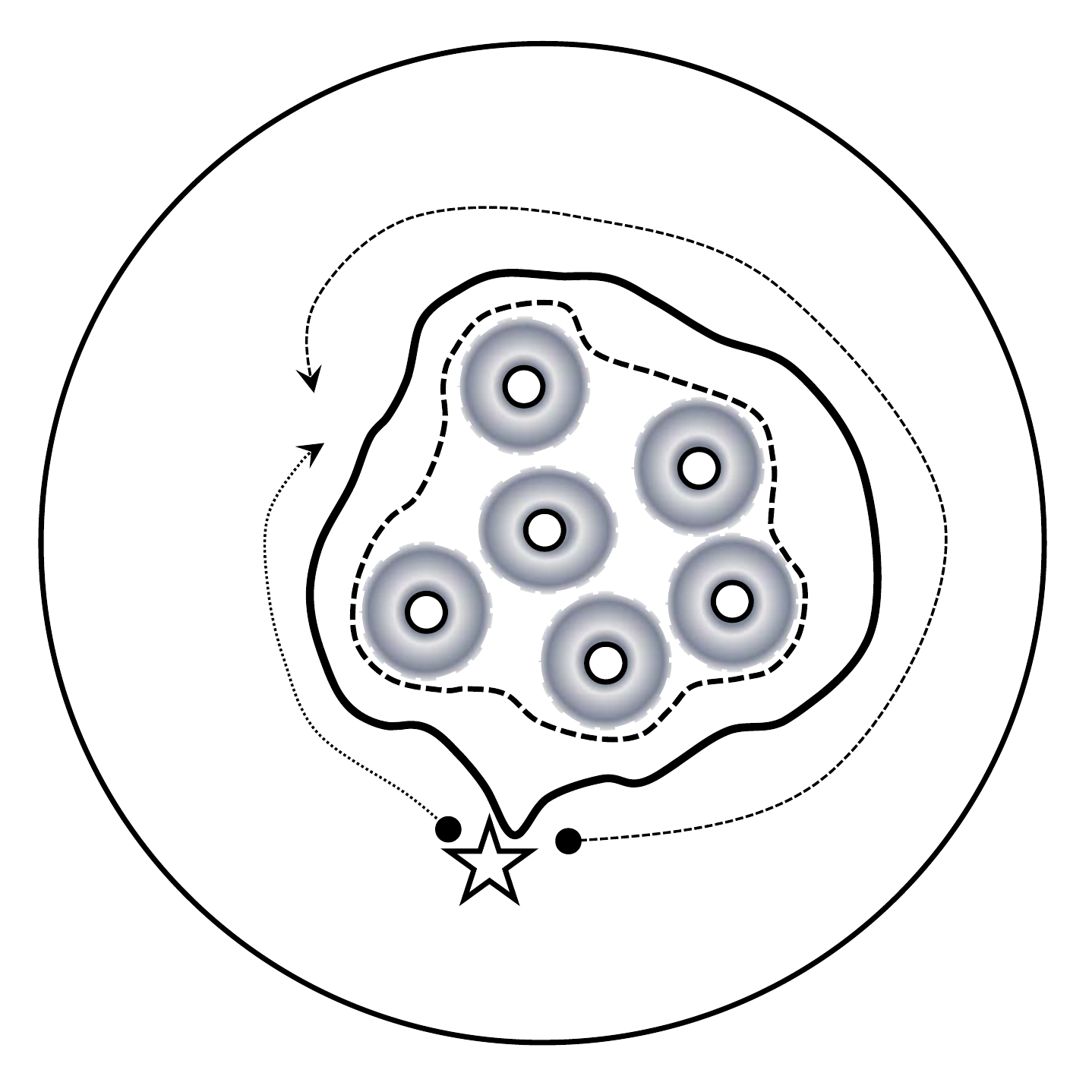}
\caption{Computation of concave hulls of point set $\mathbf P$. Scheme of the plasmodium 
interaction with $\mathbf P$: points of $\mathbf P$ are circles; diffusing repellents, 
visible as black halo's surrounding pills in Fig.~\ref{example1}ac, are shown 
by gray gradients; feeding boundary of protoplasm is marked by dotted 
line; major protoplasmic tube, which approximates $CH({\mathbf P})$, is shown by solid line.
}
\label{example1scheme}
\end{figure}

We tested feasibility of the idea in 25 experiments. All experiments were successful. In each experiment we arranged 4-8 half-pills (they represent given planar set $\mathbf P$) in a random fashion near centre of a Petri dish and inoculated an oat flake colonised by plasmodium 2-4~cm away from the set $\mathbf P$. A typical experiment is illustrated in Fig.~\ref{example1}. In 12~h after inoculation plasmodium propagates towards set $\mathbf P$ and starts enveloping the set with its body and network of protoplasmic tubes (Fig.~\ref{example1}). The plasmodium completes approximation of a shape by entirely enveloping $\mathbf P$ in next 12~h (Fig.~\ref{example1}). The plasmodium does not propagate inside configuration of pills (Fig.~\ref{example1}). Scheme of the concave hull construction is shown in Fig.~\ref{example1scheme}.

\begin{figure}
\centering
\includegraphics[width=0.45\textwidth]{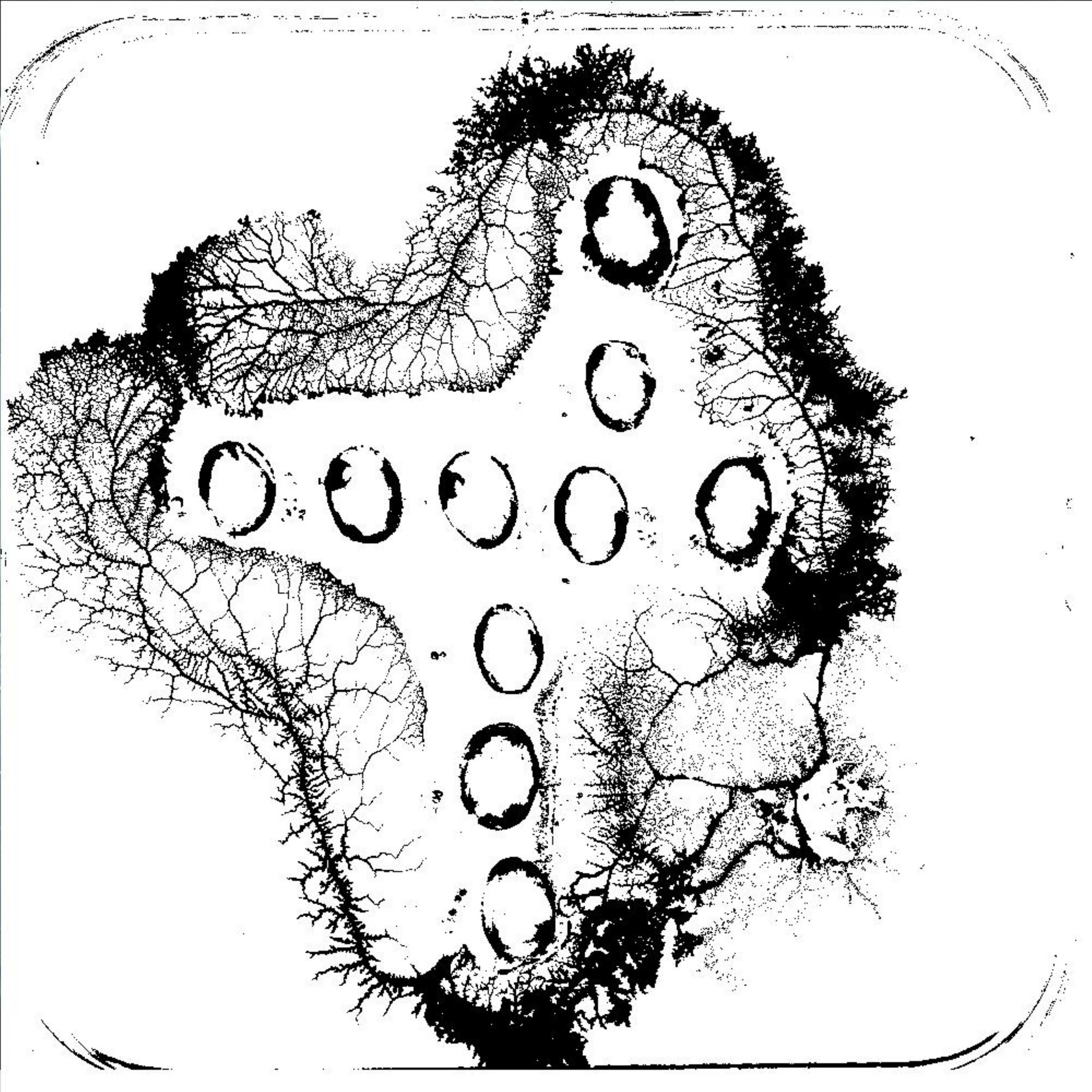}
\caption{Binarised image of plasmodium approximating concave hull of $\mathbf P$.}
\label{example2}
\end{figure}

Configuration of $\mathbf P$ in Fig.~\ref{example1}a--d favours approximation of a convex hull. If spatial configuration of points curves inwards then concave hull is approximated (Fig.~\ref{example2}). 

\begin{figure}
\centering
\includegraphics[width=0.9\textwidth]{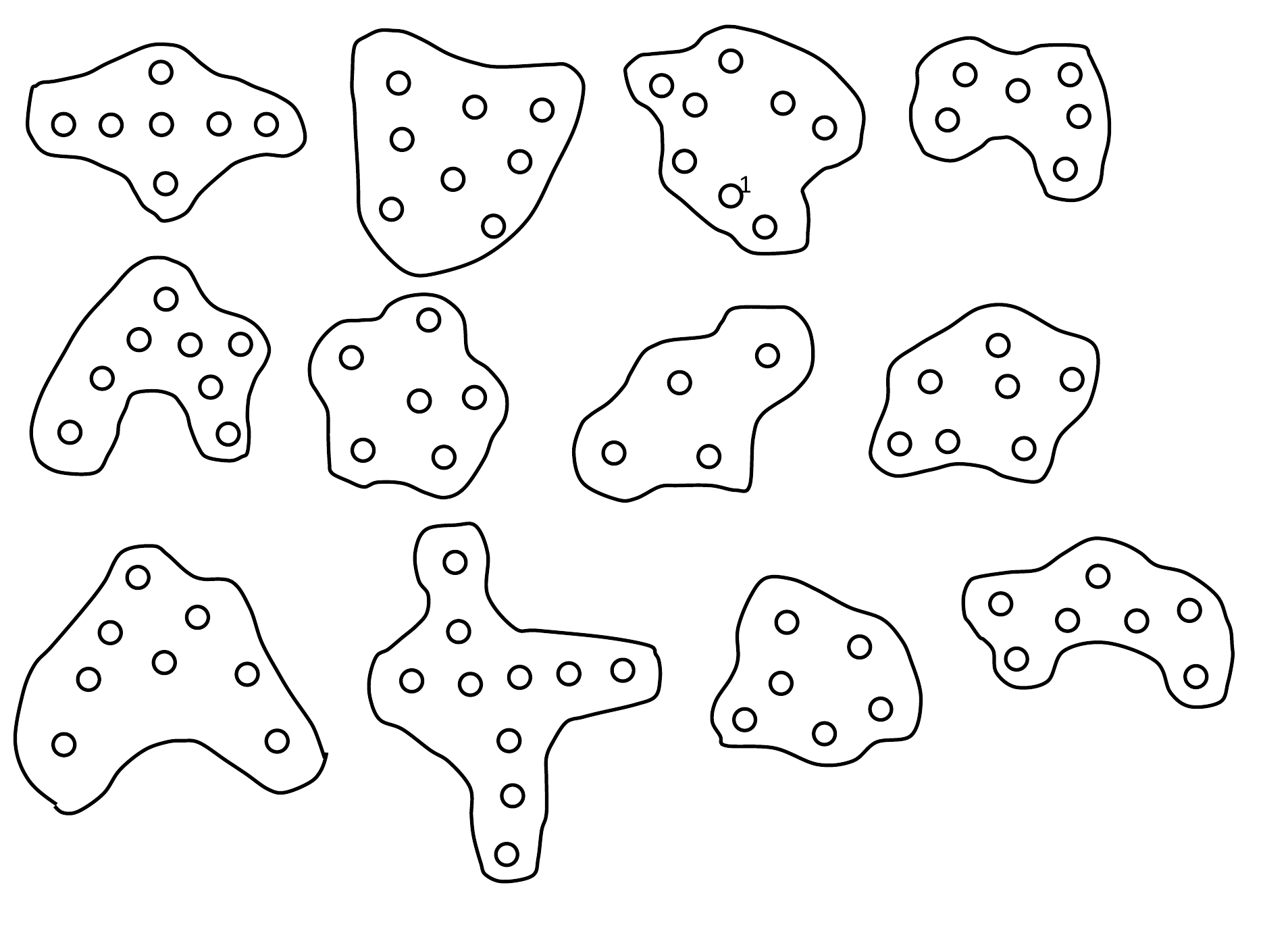}
\caption{Contours of convex and concave hulls constructed on small sets of planar points are shown. The contours are extracted in laboratory experiments with \emph{P. polycephalum}.}
\label{examplecontours}
\end{figure}

Typical hulls approximated by plasmodium of \emph{P. polycephalum} in our experiments are shown in Fig.~\ref{examplecontours}. Hulls constructed by plasmodium match their counterparts calculated by classical algorithms.

\begin{figure}
\centering
\subfigure[]{\includegraphics[width=0.45\textwidth]{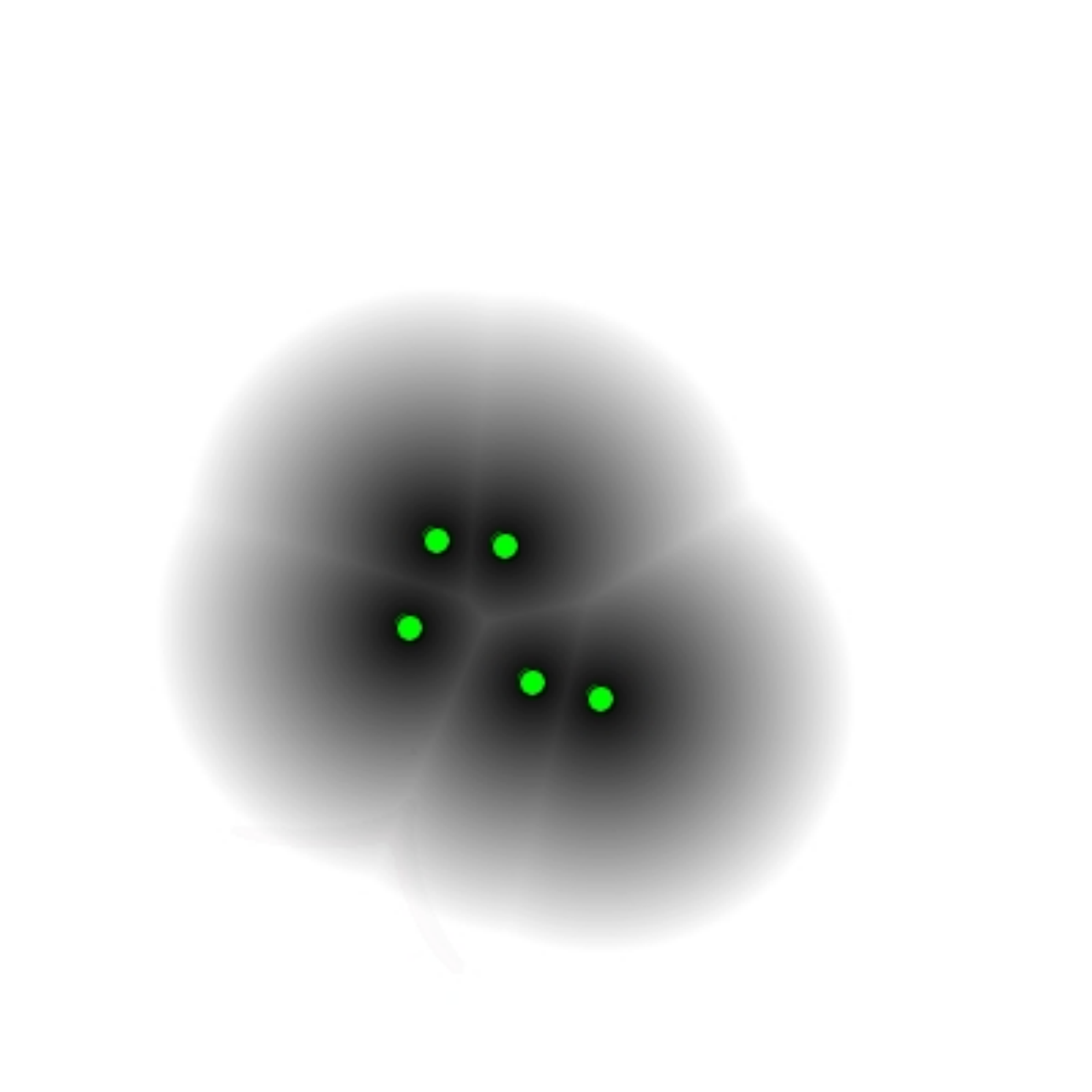}}
\subfigure[]{\includegraphics[width=0.45\textwidth]{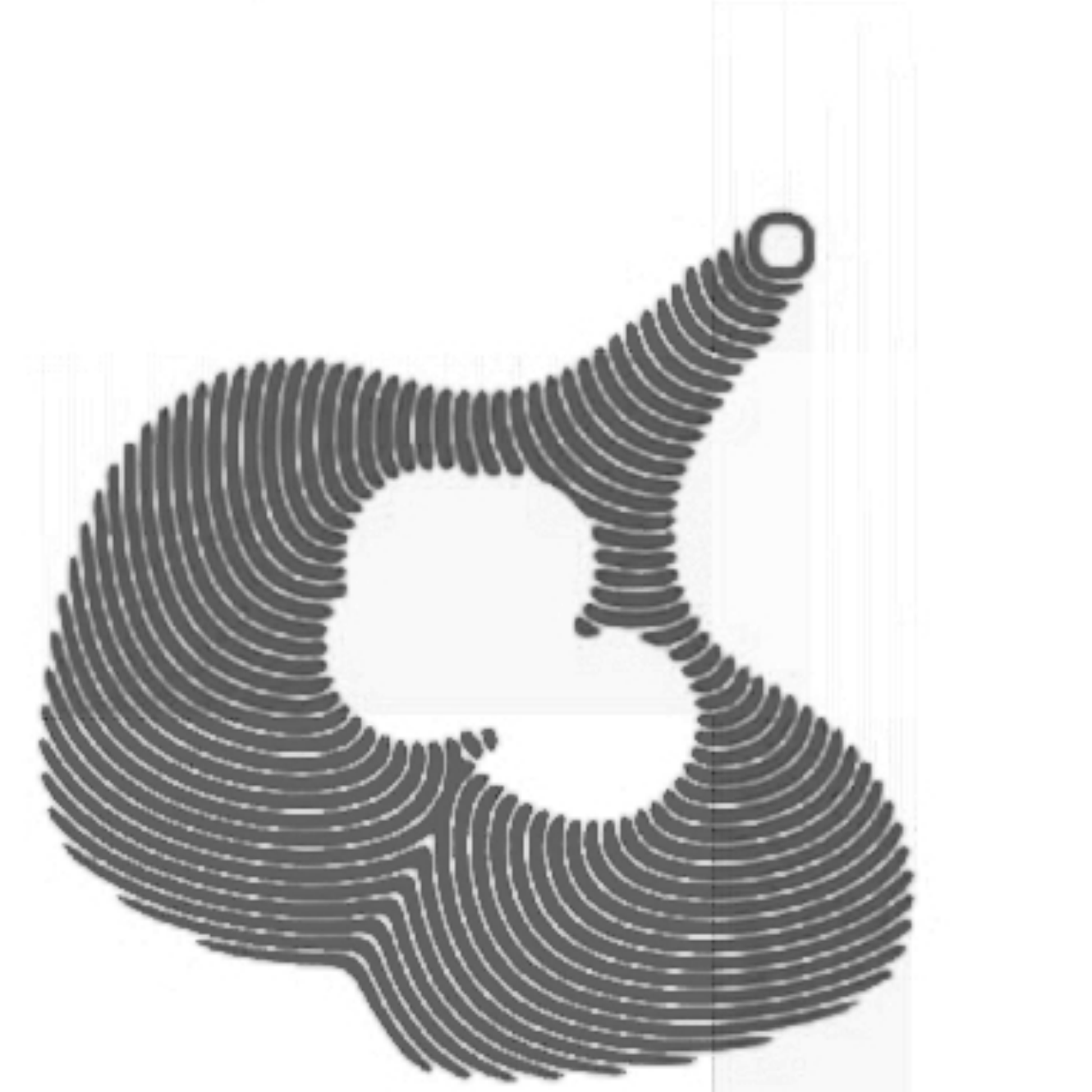}}
\subfigure[]{\includegraphics[width=0.45\textwidth]{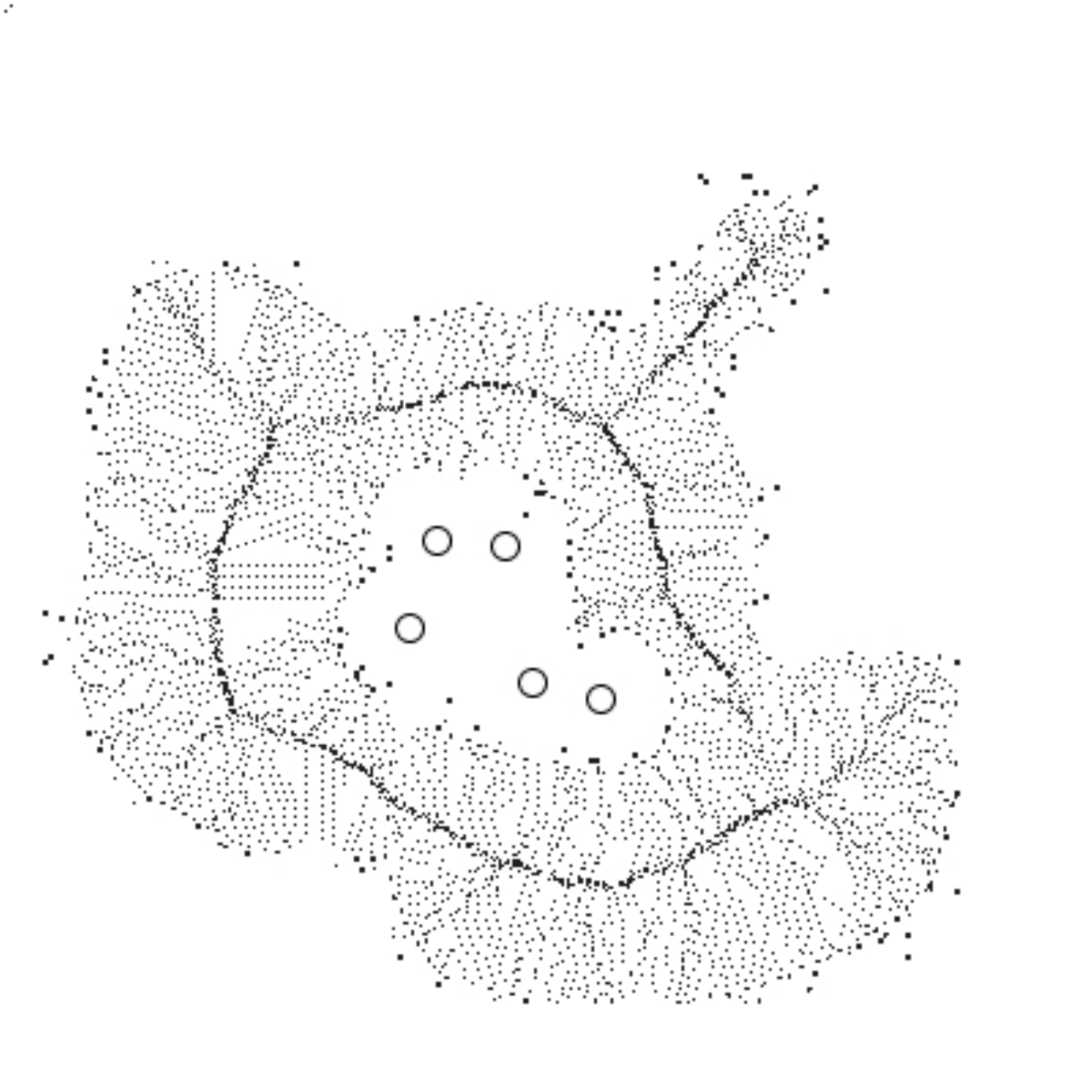}}
\caption{Simulating plasmodium approximation of convex hull of planar set $\mathbf P$ in Oregonator model. 
(a)~Gradients of repellents generated by points of $\mathbf P$. 
(b)~Time lapsed images of propagating plasmodium patterns. 
(c)~Structure of protoplasmic tubes developed.
}
\label{examplesimulation}
\end{figure}

Let us verify experimental laboratory results with computer simulation. To imitate formation of the protoplasmic tubes we store values of $u$ matrix $\bf L$, which is processed at the end of simulation. For any site $x$ and time step $t$ if $u_x>0.1$ and $L_x=0$ then $L_x=1$. Gradients are chemo-attractants used in simulation are visualised in 
Fig.~\ref{examplesimulation}a. The matrix $\bf L$ represents time lapse superposition of 
propagating wave-fronts (Fig.~\ref{examplesimulation}b). The simulation is considered completed when propagating pattern envelops $\mathbf P$ and halts any further motion.  At the end of simulation we repeatedly apply the erosion operation~\cite{adamatzky_physarummachines} (which represents  a stretch-activation effect~\cite{kamiya_1959} necessary for formation of plasmodium tubes) to $\bf L$. The resultant protoplasmic network (Fig.~\ref{examplesimulation}c)  provides a good phenomenological match for networks recorded in laboratory experiments.

\subsection{Logical gates}

It was already demonstrated that plasmodium can approximate Voronoi diagram and Delaunay triangulation, compute proximity graphs, approximate shortest path~\cite{adamatzky_physarummachines}. Experimental implementation of concave hull computing by \emph{P. polycephalum}~\cite{adamatzky_PhysarumHull} put a logical closure in our designs of real-life prototypes of Physarum machines for computational geometry.  Collision-based computing~\cite{adamatzky_CBC}  could be yet another domain of unconventional computing where the unique interaction of slime mould with herbal tablets can be explored. Previously we have demonstrated~\cite{adamatzky_gates} that due to 
inertial component of its movement active zones of \emph{P. polycephalum} implement basic logical gates in a geometrically constrained substrate. They propagate along channels cut from agar plate, one plasmodium can block and deflect propagation of another plasmodium~\cite{adamatzky_gates}. Mechanics of interaction of plasmodium with sedative tablets offer an opportunity to implement plasmodium-based logical gates in a free space, architecturelss, medium.

\begin{figure}[!tbp]
\centering
\subfigure[]{\includegraphics[scale=0.42]{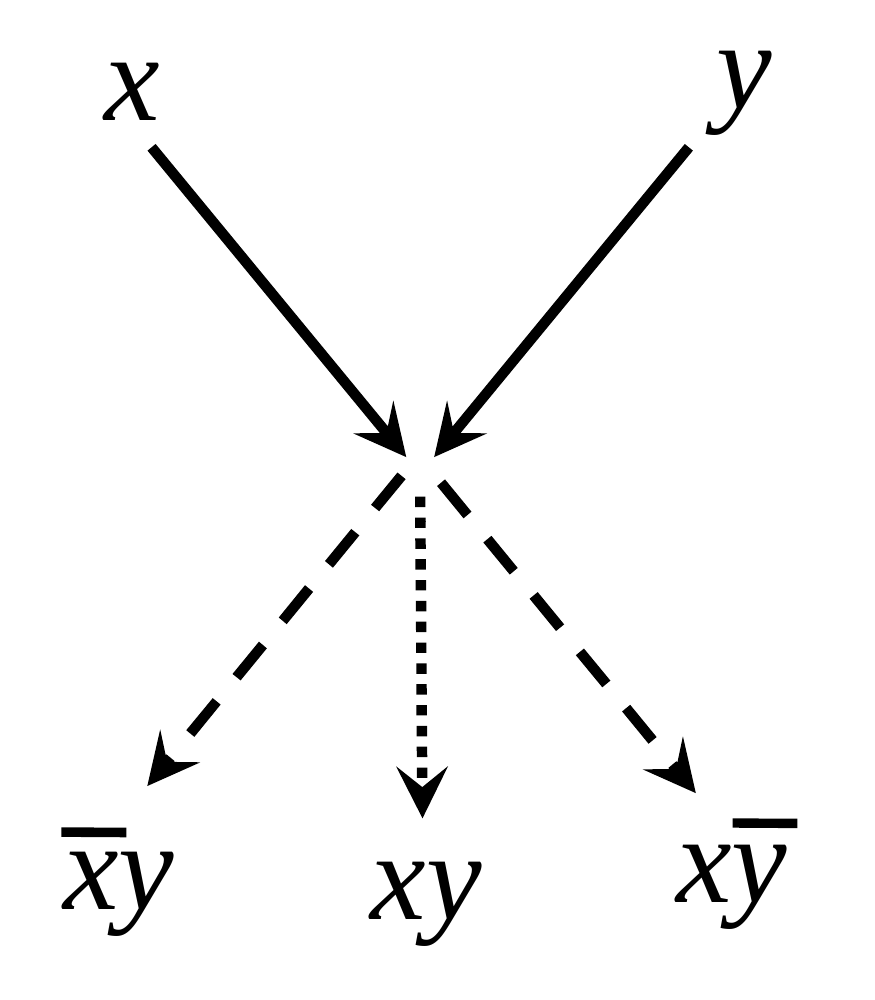}}
\subfigure[]{\includegraphics[scale=0.42]{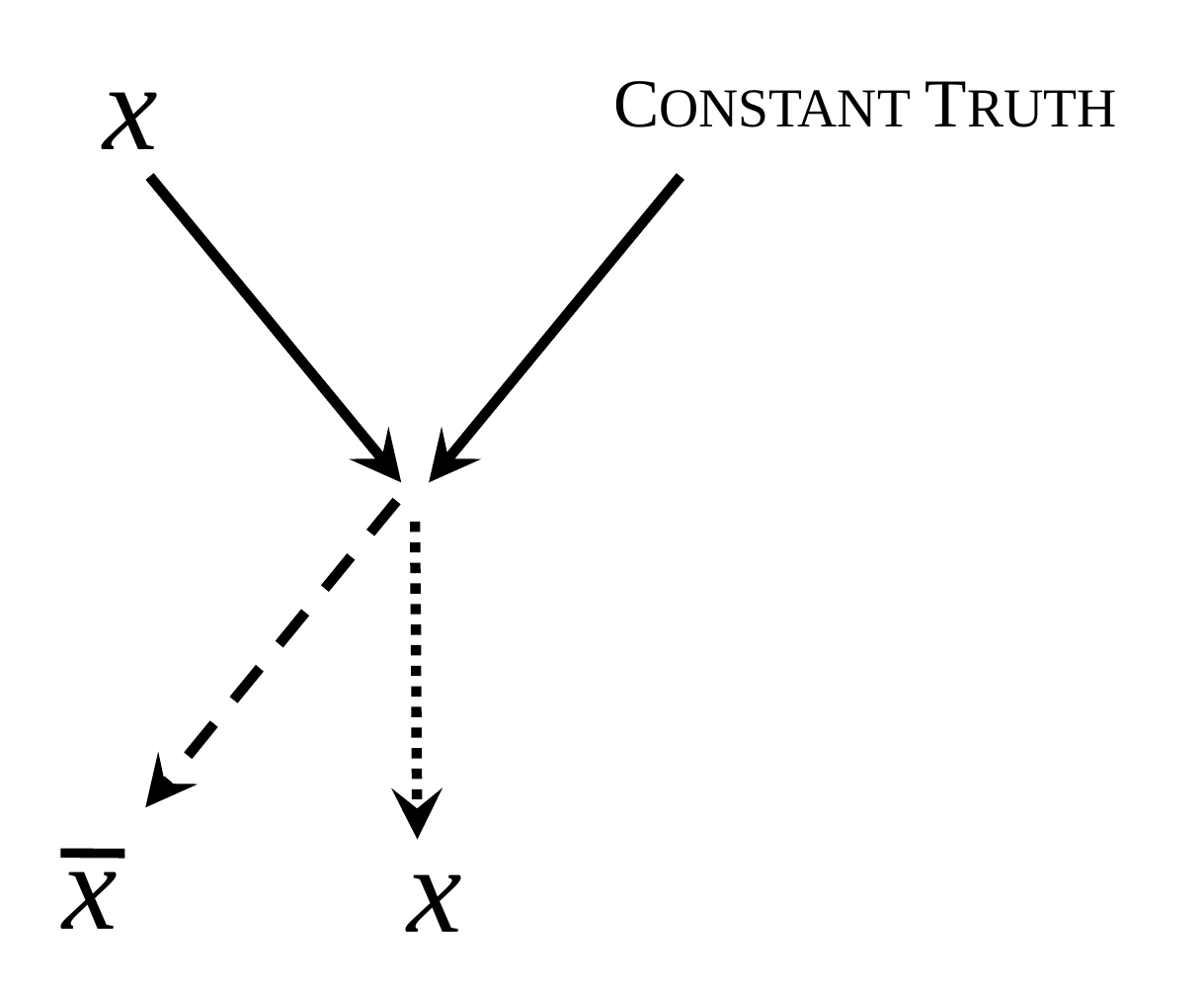}}
\subfigure[]{\includegraphics[scale=0.42]{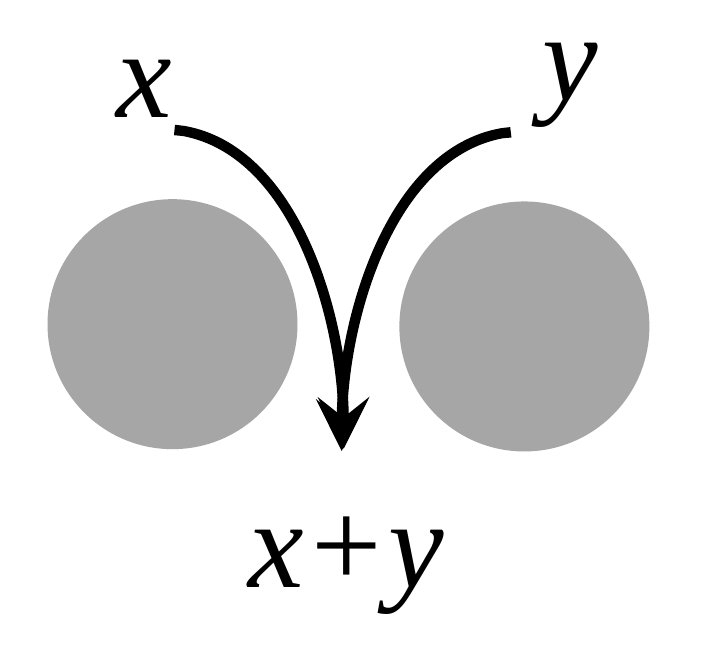}}
\caption{Scheme of collision-based gates implementable in plasmodium-tablets system.
(a)~ $\langle x, y \rangle \rightarrow \langle \overline{x}y, xy, x\overline{y} \rangle$ gate can be implemented in interaction between active zones $A$ and $C$ as illustrated in Fig.~\ref{coolfragments}de,
solid arrows are incoming trajectories of active zone $A$ (variable $x$) and active zone $C$ (variable $y$), 
dotted arrow is outgoing trajectory of active zone $D$, dashed arrows represent undisturbed trajectories of 
zones $A$ and $C$.
(b)~Negation. 
(c)~Possible implementation of conjunction. }
\label{gate}
\end{figure}

\begin{proposition}
Using herbal tablets we can implement functionally complete set of logical functions: conjuction, negation and disjunction.
\end{proposition}

In a framework of collision-based computing~\cite{adamatzky_CBC}, propagating active zones of the plasmodium, e.g. zones $A$--$D$ in Fig.~\ref{coolfragments}, are considered as `quanta' of information: a presence of a zone in a given space domain instantiates Boolean {\sc Truth} and absence of the zone in the domain Boolean {\sc False}. When two active zones collide they change their velocity vectors, e.g. deflect or merge. Post-collision trajectories of active zones 
represent Boolean values of logical operation computed in the zones' collision. 

Consider scenario shown in  Fig.~\ref{coolfragments}de. Active zone $A$ collides with active zone $C$. They fuse in new active zone $D$. If zone $A$ ($C$)  was absent then zone $C$ ($A$) would continue along its original trajectory (Fig.~\ref{gate}a). Assuming zone $A$ represents logical value of a Boolean variable $x$ and zone $C$ value of a variable  $y$ the colliding zones implement the gate $\langle x, y \rangle \rightarrow \langle \overline{x}y, xy, x\overline{y} \rangle$. Trajectory of propagation of undisturbed (i.e. not colliding with zone $A$) 
zone $C$ represents $\overline{x}y$, undisturbed trajectory of zone $A$ represents $x\overline{y}$. Zone $D$, produced in the result of the collision between $A$ and $C$, represents conjunction $xy$ of input variables $x$ and $y$. 

If we assume that input $y$ is a constant {\sc Truth} then negation gate will be as in Fig.~\ref{gate}b. 
This is a two-output gate, one output is a logical negation and other output is identity.  A construct of a disjunction gate is shown in Fig.~\ref{gate}c.  A functionally complete logical system is universal. Thus we provided a sketchy proof that a plasmodium-tablets system is a universal computing device.

All these results are rather indicative. More experiments --- in laboratory and computer simulation --- will be required to develop a 'universal programming pill', which will combine all necessary components to program advanced computation with acellular slime mould \emph{P. polycephalum} 

\section{Discussion}
\label{discussion}

\begin{figure}[!tbp]
\centering
\subfigure[]{\includegraphics[width=0.45\textwidth]{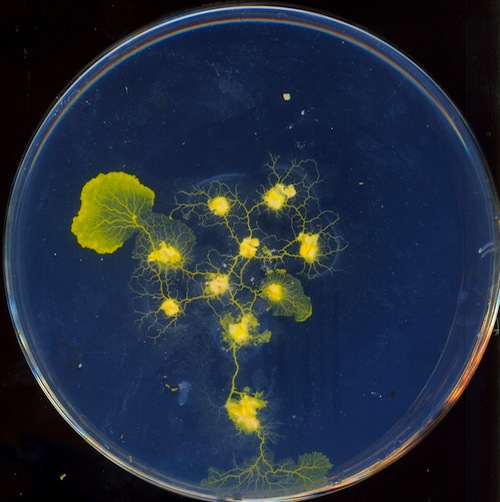}}
\subfigure[]{\includegraphics[width=0.45\textwidth]{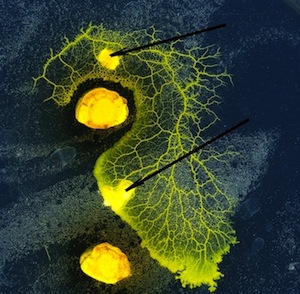}}
\caption{(a)~Plasmodium spans oat flakes by a network of protoplasmic tubes, 
(b)~plasmodium occupies oat flakes (shown by arrows) but does not come into direct contact with Kalms Tablets,
In all experiments illustrated plasmodium grows on a  non-nutrient substrate. }
\label{oats}
\end{figure}

We developed computer model imitating behaviour of slime mould \emph{P. polycephalum} in presence of herbal sedative tablets. In laboratory experiments and numerical simulation we demonstrated that the herbal tablets Nytol, Kalms Sleep and Kalms Tablets play a role of fixed attractors  and limit cycle in the space-time dynamics of the plasmodium of \emph{P. polycephalum}. The plasmodium reacts to presence of the tablets in its environment in a particular manner, drastically different from how the plasmodium reacts to conventional sources of nutrients, esp. carbohydrates. When presented with a configuration of rolled oats, or oat flakes, the plasmodium propagates towards the flakes and spans them with a network of protoplasmic tubes (Fig.~\ref{oats}a).  In a mixed configuration of oat flakes and herbal tablets, oat flakes are still used as sources of nutrients however a presence of the oat flakes does not change overall pattern of plasmodium's foraging behaviour  (Fig.~\ref{oats}b).

\begin{figure}[!tbp]
\centering
\subfigure[]{\includegraphics[width=0.45\textwidth]{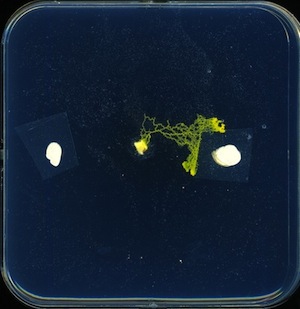}}
\subfigure[]{\includegraphics[width=0.45\textwidth]{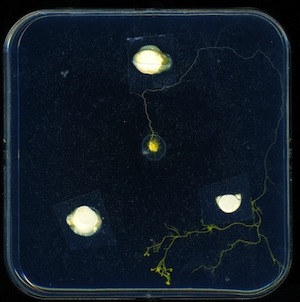}}
\subfigure[]{\includegraphics[width=0.45\textwidth]{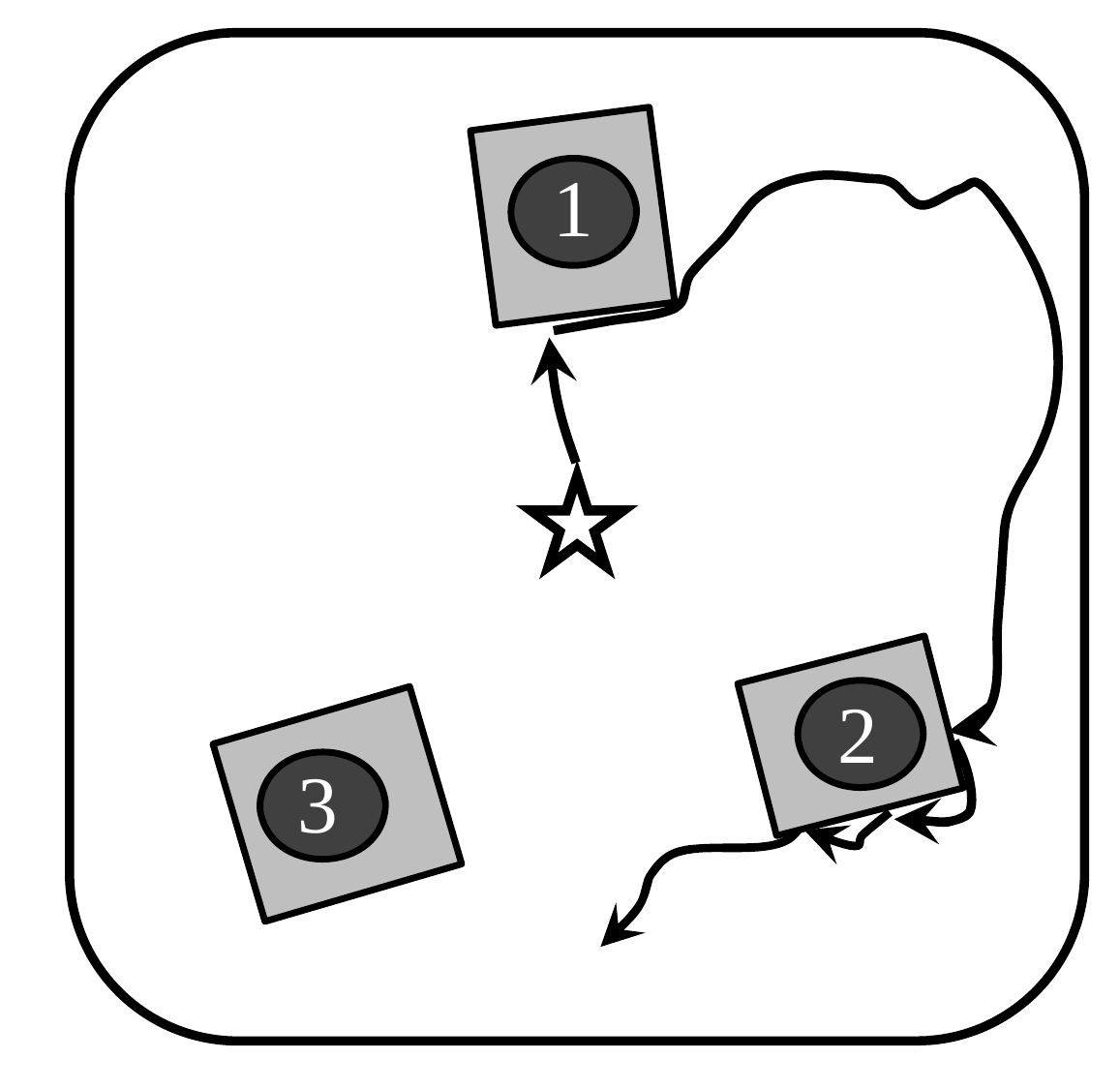}}
\caption{(a)~plasmodium is attracted to Kalms Tablets tablet placed on plastic pad (right) and does not propagate towards oat flake (left), (b)~plasmodium detects Kalms Tablets via airborne gradients of chemo-attractants and attempts to approach two tablets, (c)~scheme of the experiment (b). In all experiments illustrated plasmodium grows on a  non-nutrient substrate. }
\label{oatscde}
\end{figure} 

When given a choice between a classical source of carbohydrates, oat flakes or pasta/bread or even honey, and the herbal  tablets the plasmodium prefers herbal tablets in over 70\% of experiments~\cite{adamatzky_herbal}, see 
example in Fig.~\ref{oatscde}a. The plasmodium detects location of herbal tablets even when they are isolated 
from agar gel substrate. Thus in experiments illustrated in (Fig.~\ref{oatscde}bc) we placed tablets on polyethylene pads, which prevented diffusion of substances from tablets to agar gel. Thus plasmodium was unable to rely on gradients of chemoattractants in its substrate but only the chemo-attractants in the air. In all such experiments plasmodium successfully detects the tablets and propagates towards them. As we see in example Fig.~\ref{oatscde}bc the plasmodium detects location of tablet 1, propagates towards this tablet and tries to climb the polyethylene pad to get closer to the tablet. The plasmodium fails and then it propagates towards tablet 2 and makes few more attempts to mount the tablet's plastic pad. 

Based on our experiments  we can assume that the herbal tablets  contain at least two types of chemical substances: one substance plays a role of long-distance chemo-attractants, another substance is a short-distance repellent.
Amongst six herbal extracts present in the tablets~\cite{leaflets}, only Valerian roots (\emph{Valeriana officinalis}) 
and Hops (\emph{Humulus lupulus}) extracts are present in all three types of tablets. Thus we can propose either
Valerian or Hops is a long-distant attractant for \emph{P. polycephalum}.  To check which one, we undertook an extensive range of experiments with dried plants and found that with probability 0.73 plasmodium of \emph{P. polycephalum} prefers dried Valerian root to dried Hops leaves, see details in \cite{adamatzky_herbal}. 

What exact substance in Valerian attracts plasmodium remains unclear. 
Valerian contains hundreds of identified components including 
alkaloids ~\cite{torssell_wahlberg_1966}, volatile oils~\cite{hendriks_bruins_1980}, 
valerinol~\cite{jommi_1967}, and actinidine~\cite{johnson_wallera_1971}. Isovaleric acid and actinidine are identified in anal gland secretion of \emph{Iridomyrmex nitidiceps }ant, and isovaleric acid is considered to be a distress indicator~\cite{cavill_1982}. Possibly there is a link between the said findings, and isovaleric acid combined 
with actinidine could be considered as pheromones
of \emph{P. polycephalum} (we could cite only relevant studies conducted on cellular slime 
moulds~\cite{lewis_oday_1979, newell_1981, nader_shipley_1984}). By allowing the plasmodium to choose between Valerian and Catnip (\emph{Nepeta cataria}) we ruled out actinidine, because in all experiments  \emph{P. polycephalum} prefers Valerian to Catnip. 

Amongst components of the tablets' filling and coating only stearic acide and magnesium stearate are reportedly present in all three tablets~\cite{leaflets}. Also Kalms Sleep and Kalms Tablets tablets contain substantial (up to 160-200 mg~\cite{leaflets}) amount of sucrose. Influence of stearic acid and magnesium stearate on slime mould is not studied, and we were unable to find any published data. Sucrose would be main candidate on a role of short-distance repellents, responsible for the formation of a `no-go zone' (Fig.~\ref{singletablet}c) and, possibly, formation of feeding zone in the region with low concentration of a sucrose. Reports about influence of sugars are on the plasmodium are somewhat contradictory. Some studies suggest that plasmodium of \emph{P. polycephalum} is indifferent to sucrose, 
fructose and ribose~\cite{carlile_1970,knowles_carlile_1978} while others demonstrate that plasmodium is actually repelled by sucrose~\cite{ueda_1976}. Galactose and mannose reportedly inhibit plasmodium's motion, in situations of direct contact, yet they do not act as long-distance repellents~\cite{denbo_miller_1978}. 

Despite the fact that exact mechanism of the herbal tablets  influence on the plasmodium of \emph{P. polycephalum} remains unclear, our present findings convincingly show that the herbal tablets containing Valerian extract (and possibly Hops) could be efficiently used as control stimuli in programming and controlling behaviour of Physarum machines~\cite{adamatzky_physarummachines}. The herbal tablets is a cheap, user-friendly, and efficient alternative to existing techniques of controlling slime moulds: 
illumination-~\cite{nakagaki_yamada_1999}, thermo- \cite{tso_mansour_1975,matsumoto_1980}, and salt-based repellents~\cite{adamatzky_physarum_salt}, and carbohydrate-based 
attractants~\cite{carlile_1970,knowles_carlile_1978,dussutour_2010}.

What are potential application domains of our results? Indeed several important questions on sensitivity of plasmodium to active substances present in sedative tablets must be answered in a way of rigorous biological experiments. Exact mechanisms of slime mould's chemo-attraction are still to be discovered and plasmodium preferences to various chemical components are to be classified. In the framework of biological computing our results provide a viable alternative to existing approaches and techniques of programming experimental laboratory biological substrates. A single table can cause a wide range of behavioural patterns in slime mould and therefore --- ideally --- the whole program of slime mould based computer can be encoded in a single tablet.  Slime mould of \emph{P. polycephalum} could also play a role of an experimental laboratory model in design of amorphous soft-bodied robots and their application of substance transportation in confined environment and drug delivery inside human bodies. Our findings on control using combinations of long-distance attracts and short-distance repellents will allow for a precise navigation and control of such amorphous robots.

\end{document}